\RequirePackage{snapshot}
\documentclass{nature}

\def\ArticleTitle{Analog Iterative Machine (AIM): using light to solve quadratic optimization problems with mixed variables}
\usepackage[svgnames]{xcolor}
\usepackage[english]{babel} \usepackage{blindtext}

\usepackage{microtype}

\usepackage{amsmath,amsfonts,amsthm} \usepackage{comment}
\usepackage{endnotes}
\makeatletter
\renewcommand*\makeenmark{\hbox{\textsuperscript{\@Alph{\theenmark}}}}
\makeatother

\usepackage{booktabs}

\usepackage{lastpage}

\usepackage{graphicx} \graphicspath{{Figures/}  }

\usepackage{enumitem} \setlist{noitemsep}

\usepackage{sectsty} \allsectionsfont{\usefont{OT1}{phv}{b}{n}}

\usepackage{xargs}
\usepackage{xspace}

\usepackage[unit-optional-argument,use-xspace]{siunitx}
\usepackage[all]{foreign}

\usepackage[boxed,noline]{algorithm2e}
\usepackage{tikz}
\usetikzlibrary{trees}

\usepackage{flushend}
\usepackage{xurl}

\usepackage{geometry}

\usepackage[T1]{fontenc} \usepackage[utf8]{inputenc}

\usepackage{XCharter}

\usepackage{xcolor}

\newcommand{\emailstyle}[1]{{\footnotesize\usefont{OT1}{phv}{m}{n}\color{black}\href{mailto:#1}{#1}}}

\usepackage{lettrine} \usepackage{fix-cm}

\usepackage{xstring}

\usepackage[
  backend=biber,
  abbreviate=true,
  giveninits=true,
  isbn=false,
  url=false,
  doi=false,
  eprint=false,
  sorting=none
]{biblatex}
\renewbibmacro{in:}{}
\AtEveryBibitem{\clearfield{day}\clearfield{month}\clearfield{endday}\clearfield{endmonth}\clearlist{location}\clearlist{address}\clearfield{titleaddon}\clearfield{pages}\clearfield{language}\clearlist{editor}\clearfield{series}}

\usepackage[autostyle=true]{csquotes}

\theoremstyle{definition} 

\theoremstyle{plain}

\theoremstyle{remark} 

\usepackage{hyperref}
\hypersetup{
colorlinks=true, breaklinks=true, bookmarks=true,bookmarksnumbered,
urlcolor=brown, linkcolor=RoyalBlue, citecolor=green, pdftitle={\ArticleTitle}, pdfauthor={Microsoft \textcopyright 2022}, pdfsubject={}, pdfkeywords={}, pdfcreator={}, pdfproducer={} }

\makeatletter
\newcommand\Autoref[1]{\@first@ref#1,@}
\def\@throw@dot#1.#2@{#1}\def\@set@refname#1{\edef\@tmp{\getrefbykeydefault{#1}{anchor}{}}\xdef\@tmp{\expandafter\@throw@dot\@tmp.@}\ltx@IfUndefined{\@tmp autorefnameplural}{\def\@refname{\@nameuse{\@tmp autorefname}s}}{\def\@refname{\@nameuse{\@tmp autorefnameplural}}}}
\def\@first@ref#1,#2{\ifx#2@\autoref{#1}\let\@nextref\@gobble \else \@set@refname{#1}\@refname~\ref{#1}\let\@nextref\@next@ref \fi \@nextref#2}
\def\@next@ref#1,#2{\ifx#2@ and~\ref{#1}\let\@nextref\@gobble \else,~\ref{#1}\fi \@nextref#2}
\makeatother

\usepackage[nameinlink]{cleveref}
\usepackage{setspace}

\usepackage[all]{nowidow}
\usepackage{setspace}

\usepackage{amsfonts}
\usepackage{amssymb}
\usepackage{amsmath}

\usepackage{float}
\usepackage{placeins}
\usepackage{graphicx}

\usepackage[export]{adjustbox}

\usepackage[disable]{endfloat}

\usepackage[normal,bf,labelformat=empty]{caption}

\newcounter{mybodyfigure}
\newcounter{myedfigure}

\newcommand{\beginbodyfigures}{\renewcommand{\thefigure}{{\themybodyfigure}}}

\newcommand{\beginedfigures}{\renewcommand{\thefigure}{{\themyedfigure}}}

\newcommand{\stepbodyfigure}{\refstepcounter{mybodyfigure}}

\DeclareRobustCommand{\bodyfigure}[1]{\stepbodyfigure\label{#1}{\themybodyfigure}}

\newcommand{\bodyfigurelabel}[1]{\bf{Figure \bodyfigure{#1}:}}

\makeatletter
\g@addto@macro\caption@prepareslc{\renewcommand{\stepbodyfigure}{\caption@l@stepcounter{mybodyfigure}}}
\makeatother

\newcommand{\stepedfigure}{\refstepcounter{myedfigure}}

\DeclareRobustCommand{\edfigure}[1]{\stepedfigure\label{#1}{\themyedfigure}}

\newcommand{\edfigurelabel}[1]{\bf{Extended Data Figure \edfigure{#1}:}}

\makeatletter
\g@addto@macro\caption@prepareslc{\renewcommand{\stepedfigure}{\caption@l@stepcounter{myedfigure}}}
\makeatother

\usepackage{hyperref}
\usepackage{verbatim}

\usepackage{ulem} \usepackage{rotating} \usepackage{soul}

\usepackage{color}

\usepackage[colorinlistoftodos]{todonotes}

\makeatletter
\newsavebox\myboxA
\newsavebox\myboxB
\newlength\mylenA

\newcommand*\xoverline[2][0.75]{\sbox{\myboxA}{$\m@th#2$}\setbox\myboxB\null \ht\myboxB=\ht\myboxA \dp\myboxB=\dp\myboxA \wd\myboxB=#1\wd\myboxA \sbox\myboxB{$\m@th\overline{\copy\myboxB}$}\setlength\mylenA{\the\wd\myboxA}\addtolength\mylenA{-\the\wd\myboxB}\ifdim\wd\myboxB<\wd\myboxA \rlap{\hskip 0.5\mylenA\usebox\myboxB}{\usebox\myboxA}\else
        \hskip -0.5\mylenA\rlap{\usebox\myboxA}{\hskip 0.5\mylenA\usebox\myboxB}\fi}
\makeatother

\title{\ArticleTitle}

\author{Kirill P. Kalinin$^1$, George Mourgias-Alexandris$^1$, Hitesh Ballani$^1$,
  Natalia G. Berloff$^{2,1}$,
  James Clegg$^1$,
	Daniel Cletheroe$^1$,Christos Gkantsidis$^{1,*}$,
  Istvan Haller$^1$,
  Vassily Lyutsarev$^1$, Francesca Parmigiani$^{1,*}$, Lucinda Pickup$^1$ and Antony Rowstron$^1$
}

\graphicspath{ {../Artwork/} }
\def\AIM{\emph{AIM}\xspace}

\def\QUMO{\emph{QUMO}\xspace}
\def\QUBO{\emph{QUBO}\xspace}

\makeatletter
\DeclareRobustCommand\onedot{\futurelet\@let@token\@onedot}
\def\@onedot{\ifx\@let@token.\else.\null\fi\xspace}

\makeatother

\bibliography{aim-main.bib}

\begin{document}\maketitle

\begin{affiliations}
  \item Microsoft Research, 21 Station Road, Cambridge CB1 2FB, UK
  \item Department of Applied Mathematics and Theoretical Physics, University of Cambridge, Cambridge CB3 OWA, UK
  \item[] $^{*}$\emailstyle{project-aim-contact@microsoft.com}
\end{affiliations}

\begin{abstract}

Solving optimization problems is challenging for existing digital
computers and even for future quantum hardware.
The practical importance of diverse problems, from
healthcare to financial optimization, has driven the emergence of specialised hardware
over the past decade.
However, their support for problems with only binary variables severely restricts the
scope of practical problems that can be efficiently embedded.
We build analog iterative machine (\AIM), the first instance of an opto-electronic solver
that natively implements a wider class of quadratic
unconstrained mixed optimization (\QUMO) problems and supports \emph{all-to-all}
connectivity of both continuous and binary variables.
Beyond synthetic 7-bit problems at small-scale, \AIM\ solves the financial transaction settlement problem
entirely in analog domain with higher accuracy than quantum
hardware and at room temperature.
With compute-in-memory operation and spatial-division multiplexed representation
of variables, \AIM's design paves the path to chip-scale architecture with 100 times
speed-up per unit-power over the latest GPUs for solving problems with 10,000 variables.
The robustness of the \AIM\ algorithm at such scale is further demonstrated
by comparing it with commercial production solvers across multiple
benchmarks, where for several problems we report new best solutions.
By combining the superior \QUMO\ abstraction, sophisticated gradient
descent methods inspired by machine learning, and commodity hardware, \AIM\ introduces a
novel platform with a step change in expressiveness, performance, and scalability,
for optimization in the post-Moore's law era.

\end{abstract}

 \beginbodyfigures

Optimization is a journey, requiring identification of a task as an optimization problem,
creativity in formulating it mathematically, and inventiveness in solving it. Such problems are deeply ingrained in almost every industry today, from operations
research and manufacturing, finance and engineering, to healthcare and transportation. We create an analog optimization machine that, by combining mathematical insights with
algorithmic and hardware advances,
could offer a crucial inflection point in the optimization journey.

Optimization workloads and machine learning applications are commonly accelerated with clusters of
cloud-based digital chips such as graphical processing units (GPUs). However, the amount of required processing power is
beyond traditional digital hardware that is already plateauing. Even quantum hardware cannot help with optimization at practical scales despite the
promise of significant acceleration for other applications fields
\autocite{PRX/2021/Babbush:QuadraticSpeedups}. As a consequence, several unconventional
hardware platforms for solving optimization problems have been proposed over the
past decades. Prominent examples include
optical parametric oscillators \autocite{Marandi2014NetworkOT,McMahon2016,Inagaki2016}, memristors \autocite{Cai2020}, polaritons \autocite{berloff2017realizing,kalinin2020polaritonic},
coupled lasers \autocite{pal2020rapid}, and others
\autocite{babaeian2019single,parto2020realizing,pierangeli2019large,roques2020heuristic}.
Many of these approaches resort to hybrid architecture, when digital electronics is fused with
the unconventional hardware to achieve scalability. Such hybrid approaches sacrifice their
potential speed-up substantially.
In particular, the speed-up per watt is compromised due to analog-to-digital signal conversions.
For the time-multiplexed systems, where each variable is assigned a certain time slot within a
single source signal, further speed-up reduction is caused by sequential processing of problem
variables. These limitations motivate machines with spatially-multiplexed
variables, which enable high-throughput parallel operations for
spatial differentiation \autocite{abdol2015analog}, integration
\autocite{golovastikov2015spatial}, and solving differential equations \autocite{abdol2017dielectric}.
For chip-scale platforms, the spatially-multiplexed systems promise significant speed-ups with
a recent demonstration of two-variable machine \autocite{okawachi2020demonstration}.

Beyond the hardware limitations, most unconventional hardware optimizers, starting with pioneering
Hopfield networks \autocite{PNAS/1982/Hopfield:Network,hopfield1985neural},
target optimization tasks described as the quadratic unconstrained binary
optimization (\QUBO) problem. The \QUBO
abstraction is a poor fit for many real-world problems \autocite{DARPA/2021:QuICC}.
In theory, an accurate solver for any $\mathbb{NP}$-hard problem, such as the \QUBO model, can
solve all $\mathbb{NP}$ problems with at most polynomial overhead in number of variables, which is deemed
to be negligible in algorithmic sense.
In practice, the polynomial overhead may lead to significant increase in problem sizes. The
overhead associated with limited hardware connectivity often results in additional substantial
inflation in problem sizes \autocite{Dwave:Web}. Together, the mapping and limited connectivity overheads prevent
the hardware solvers from tackling industrially-relevant applications at scale.

In the emerging realm of analog computation, where complex mathematical operations are performed
at extremely high speed in parallel, we present the analog iterative machine (\AIM) that
innovates simultaneously across three dimensions --- problem abstraction, algorithmic design, and
analog hardware architecture --- for unconventional computing paradigms.

At the abstraction level, we introduce a quadratic unconstrained mixed
optimization (\QUMO), which allows for both binary and continuous
variables. The \QUMO\ abstraction generalises the \QUBO\ formulation
and offers a more natural language to express quadratic optimization problems with linear
inequality constraints, which results in a more
efficient mapping for many applications including transaction settlement
\autocite{braine2021quantum} and Markowitz portfolio optimization
\autocite{Beasley/2013:PortfolioOptimisationSurvey} problems in financial domain,
crystallographic texture approximation problems in chemistry \autocite{Materialia/2006/Bohlke:Crystallographic},
and compressed sensing in healthcare \autocite{donoho2006compressed}.
\begin{figure}[t]
  \caption{\bodyfigurelabel{fig:intro}
    \textbf{\QUMO\ abstraction in analogue hardware.} (a) The schematic of the opto-electronic
    system in which the iterative update rule for
    the gradient of the objective function, annealing, and momentum terms are implemented in
    analogue hardware. The problem variables are encoded in the signal intensities, namely light
    intensities and electrical currents, and the problem
    input is represented with optical modulators.
  (b) The optimization problems with
  quadratic objective and linear inequality constraints can be efficiently represented within
  the \QUMO\ abstraction by introducing one additional continuous variable per constraint.}\centering
  \includegraphics[width=17.2cm]{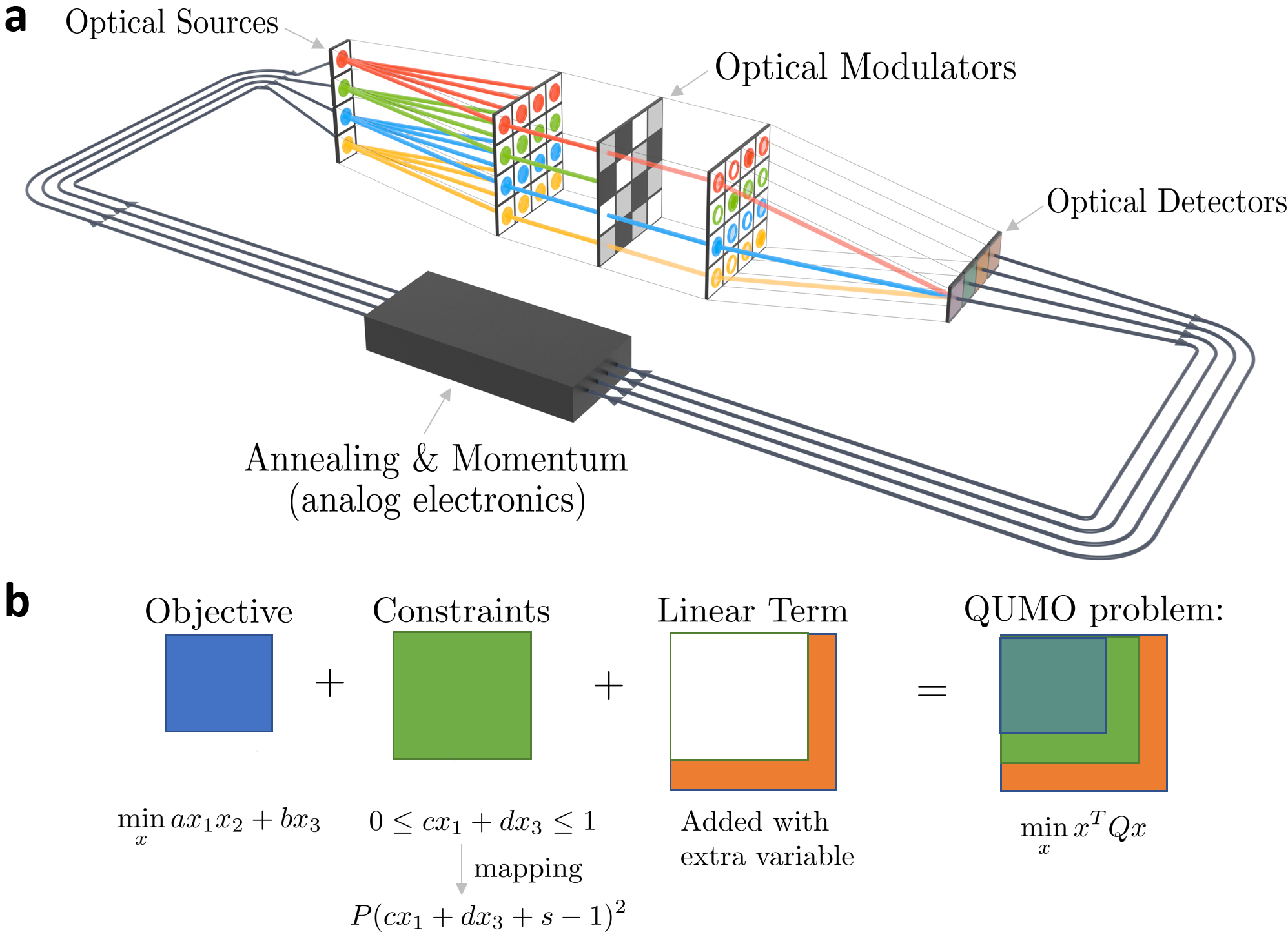}
\end{figure}

At the algorithmic level, we introduce an advanced gradient descent
search algorithm, inspired by well-known machine learning approaches. To overcome limitations of previous hardware approaches, whose methodology is often based on describing physical system
behaviour and envisioning an improved performance of hardware thanks
to natural internal processes, we focus on co-designing a highly
performing algorithm and hardware from the beginning. The \AIM\ algorithm is an iterative approach that builds on three core
components: gradient of the objective function, annealing, and
momentum acceleration. For the \QUMO\ abstraction, the gradient is represented by a
matrix-vector product, while annealing and momentum techniques are
essential for achieving advantageous performance. All three techniques
have been chosen and designed to be amenable to efficient hardware
acceleration.

At the hardware level, we build the first $7$-variable instance of the
\AIM\ solver based on discrete optical components. As we show schematically in \autoref{fig:intro}(a),
fast matrix-vector multiplication is realized using optics by representing variables
as analog signal intensities, i.e., light intensities and currents, and by encoding
the optimization problem as a matrix with spatial optical modulation technologies.
Such design allows \AIM\ to natively support all-to-all pairwise connectivity between the variables.
The annealing and momentum techniques are well-suited for implementing in electronic domain.
We report fully hardware-measured performance results for a range of \QUMO\ problems
at $7$-bit precision. \AIM\ solves all problems with an average success rate over $87\%$, which is in agreement
with the corresponding simulations.
As an illustrative example of a real-world \QUMO problem, that is also representative of
a broader class of industrial optimization problems,
we consider the transaction settlement problem and demonstrate
that \AIM\ outperforms state-of-the-art quantum hardware \autocite{QE/Lee/2021:MBOforFinance}.

Three crucial design elements of \AIM\ pave the path to new architectures with a
potential $100$ times speed-up over state-of-the-art digital solvers for the same power at
$10,000$ variable scale. First, the data transfer between digital and analog domains is fully eliminated
until convergence to a solution is achieved. Second, there is no separation between compute and memory since the variables are
computed as light traverses through the optical modulation matrix. Finally, \AIM\ leverages the inherent speed and parallelism of
optics and electronics to represent each problem variable as an
independent analog signal, thus eschewing the use of time-multiplexing of variables
as a scaling technique.
We envision that scaling to ten thousand variables
is key to addressing real-world problems at scales of
practical importance, and outline how the \AIM\ architecture and technology choices open
the way for such scalability at low cost.

To evaluate the \AIM algorithm on problem sizes of interest, we utilize
a GPU-based simulation of \AIM for an extensive collection of \QUBO\ and \QUMO\ problems
including G-Set \autocite{Gset:Web}, Wishart \autocite{wishart_paper},
Tile-3D \autocite{tile3d_paper}, RCDP \autocite{rcdp_paper}, and
QPLIB \autocite{QPLib:Paper} benchmarks. The \QUBO\ problems are included to evaluate the \AIM\ algorithm relative to the
production-grade physics-inspired heuristics. Across all benchmarks, the simulated \AIM\
demonstrates competitive and often superior solution quality, compared
to state-of-the-art heuristic approaches \autocite{QIOservice:Web}, namely simulated annealing
and parallel tempering, and commercial optimization package, i.e. Gurobi \autocite{Gurobi:Web}.
Moreover, the simulated \AIM\ algorithm sets
new state-of-the-art results on two \QUMO and two \QUBO problems from QPLIB benchmark which
translate to three orders of magnitude speed-up. In doing so, the \AIM algorithm demonstrates
a leap forward in the capabilities of quadratic optimization algorithms.

 \section*{QUMO}

Optimization problems are nonlinear by their nature and, ideally, algorithms should
be designed to exploit and benefit from the original problem formulation. In practice, this approach requires a substantial expert knowledge for each class of
optimization problems, resulting in their reduction to a common abstraction for which
solvers exist. Given the long history of linear programming development, optimization problems are usually
linearised in standard optimization packages that exploit the simplex and interior point
methods. On one side, the linearisation approach allows one to treat a broad range of problems equally. On the other side, solving the problem in the form closest to the original formulation is
expected to be the most efficient  \autocite{valiante2021computational}.

The \QUMO abstraction represents a wide class of combinatorial nonlinear optimization
problems and is formulated as the minimisation of an objective function $F(\bf{x})$:
\begin{equation}
    \underset{\bf x}{\text{Minimize}} \ F({\bf x}) = \underset{\bf x}{\text{Minimize}} \ -
    \frac{1}{2} {\bf x}^T Q {\bf x} - {\bf b}^T {\bf x}
    \label{eq_qumo}
\end{equation}
where the vector $\bf{x}$ includes binary and continuous variables, $\cdot^T$ is
the transpose operator, and the information about the optimization problem is encoded in the
weight matrix $Q$ and the constant vector $\bf{b}$. Without loss of generality, one could consider the $\{0, 1\}$ values for binary and
the $[0, 1]$ interval for continuous variables. The solution to the \QUMO\ model is the assignment of the variables $\bf{x}$ that minimises
the expression in \autoref{eq_qumo}. In contrast, the components of $\bf{x}$ are binary variables in \autoref{eq_qumo} for the
\QUBO\ framework.

Besides being nonlinear, most optimization problems are constrained. A problem with linear inequality constraints exemplifies the superior expressiveness
of \QUMO\ abstraction over the standard \QUBO\ representation. As illustrated schematically in \autoref{fig:intro}(b),
only one additional continuous variable, typically
referred to as slack variable, is required for mapping one inequality constraint to the
\QUMO\ abstraction with a penalty method.
The resulting matrix
of weights $Q$ could also incorporate all the objective linear terms with an addition
of the extra binary variable.
In contrast, the \QUBO\ model suffers from a large mapping overhead: $10$ to $100$ binary variables are needed to represent a single constraint
with either binary or unary encoding (see Suppl. material for details). In addition, the smaller problem size implies that a better solution quality can be achieved
under the same time constraints as a smaller variable space needs to be explored.
We note that the \QUBO model is equivalent to the well-known Ising and maximum cut models
(see Suppl. Mat.).

\section*{\AIM algorithm}

Optimization techniques may be classified into derivate-free methods
and algorithms exploiting information about the gradient of the objective function,
i.e., gradient-based methods.
The \AIM\ algorithm is the advanced gradient descent method, represented by the following
iterative update rule:
\begin{equation}
    {\bf x}_{t+1} = {{\bf x}_t} + \Delta t \left[ -
    \alpha \nabla F({\bf f}_{\rm nonlinear}({\bf x}_t)) -
    \beta(t) {\bf x}_t +
    \gamma ({\bf x}_t - {\bf x}_{t - 1}) \right],
    \label{eq:aim_iterative_update_rule}
\end{equation}
where ${\bf x}_t$ is the continuous real-valued state vector at time
iteration $t$, $\Delta t$ is the positive number known as the time
step (or learning rate), $\nabla F$ is the gradient of the objective
function, ${\bf f}_{\rm nonlinear}(\cdot)$ is the elementwise nonlinear
function that projects the variables on binary or continuous range of values, $\alpha$ is
the objective scaling parameter, $\beta(t)$ is the annealing schedule,
and $\gamma$ is the momentum parameter. The term involving $\alpha$ on its own represents the simplest gradient descent
method, known as the steepest descent, which modifies the system
variables ${\bf x}_t$ along the direction of the negative gradient thereby
decreasing the objective. With an addition of the last term in \autoref{eq:aim_iterative_update_rule},
the steepest descent is generalised to
an accelerated gradient-based approach that is known as the heavy ball
method \autocite{Polyak/64:SpeedingUp}. In continuous time case, the momentum term corresponds to the dynamics
of the second-order differential equation, which distinguishes the \AIM\ algorithm from the first-order methods
similar to Hopfield networks \autocite{PNAS/1982/Hopfield:Network,hopfield1985neural}. The intuition behind the momentum-based methods is simple: if one assumes that the variables ${\bf x}_t$ represent coordinates of
particles, then the momentum parameter is equivalent to the mass of
particles moving through a viscous medium in a conservative force
field \autocite{qian1999momentum}. The annealing schedule $\beta(t)$ makes the system non-conservative:
it characterises the system dissipation rate and controls how much the
amplitude of the state ${\bf x}_t$ is reduced at each time iteration $t$. The \AIM\ algorithm is of a general kind and can be applied to any
objective function $F({\bf x})$, although in this study we will consider it for solving \QUMO\ problems
described by the \autoref{eq_qumo}.

\begin{figure}
\caption{\bodyfigurelabel{fig:algorithmintuition}
  \textbf{Analog Iterative Machine (\AIM) solver.}
    (a-c) The operational principles of the \AIM\ algorithm are
      depicted schematically for a single combination of parameters.
    (a) The evolution of the objective landscape is shown for time
      iterations $t \in [0, T]$. The initially flattened landscape
      facilitates exploration of the multidimensional
      variable space and eventually returns to its original form,
      when exploitation occurs and the algorithm converges to the
      minimum of the objective.
    (b) The annealing term is characterized by $\beta(t)$ that decreases linearly over time,
      ensuring exploration and exploitation stages of the algorithm.
    (c) The better objective values are generally obtained towards
      the final time iteration $T$ as the contribution of the
      objective term $\alpha$ increases relatively to $\beta(t)$-term
      in \autoref{eq:aim_iterative_update_rule}.
    (d) The two phases of the \AIM\ solver are illustrated. During
      the `exploration phase', the \AIM\ algorithm is simulated for a
      large number of parameter combinations $(\alpha_0, \beta_0)$
      with small number of time iterations and samples per each
      combination. During the `deep search' phase, the parameter pairs
      $(\alpha_0^*, \beta_0^*)$, which produce relatively better
      objective values during exploration phase, are selected and the
      \AIM\ algorithm is simulated for a large number of time
      iterations and samples per each pair of parameters.
}
    \bigskip
    \centerline{\includegraphics[width=0.6\textwidth]{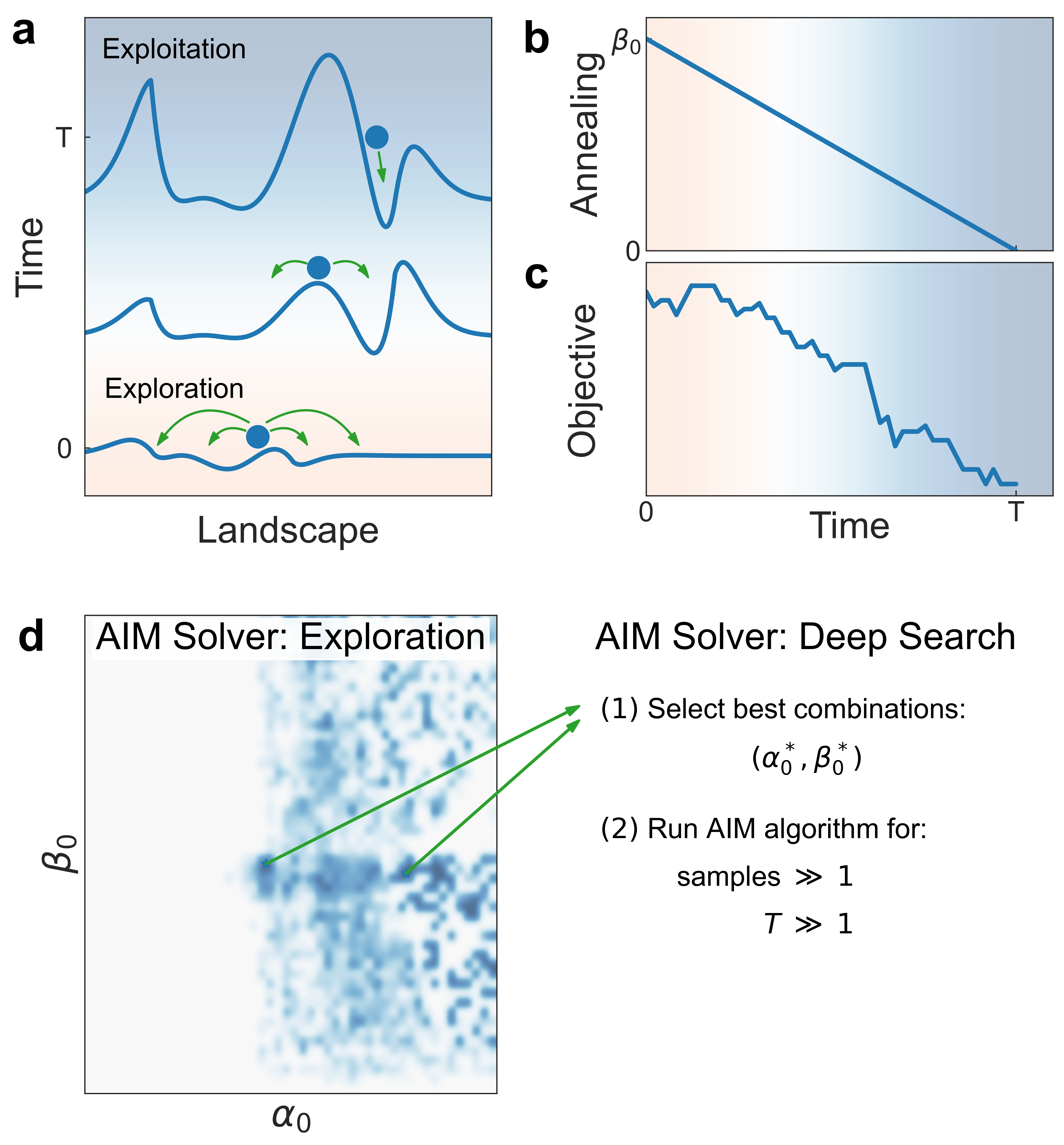}}
\end{figure}

The distinction of the \AIM\ algorithm is the simultaneous inclusion of both
momentum and annealing terms, which dramatically improve
the performance of the standard steepest gradient descent method on
nonconvex optimization problems. As shown schematically in \autoref{fig:algorithmintuition}(a),
the annealing schedule $\beta(t)$ enhances exploration over
multi-dimensional objective function space: it suppresses the contribution from the gradient of the objective
function during initial time iterations, thereby flattening the
objective profile. According to the above physical interpretation of the momentum term,
the massive particles accelerate their motion in long and narrow
valleys, improving the convergence of the iterative approach to
minima and providing mechanism for escaping from local
minima. From a numerical perspective, the presence of momentum term increases
the range of time step values for which the system
converges \autocite{qian1999momentum}. In machine learning, the momentum-based approaches are known to
greatly improve the speed of training, while the annealing schedule is
reminiscent of slowly decaying weights and could be seen as the
regularisation technique. The variations of the \AIM\ algorithm for the Nesterov momentum and
adaptive momentum methods are discussed in Supplementary materials.

The general tendency of the \AIM\ algorithm to achieve better objective
values towards the final time iteration $T$ is ensured by
relatively increasing contributions from the gradient of the objective
with respect to the annealing and momentum terms. The annealing term may be time-dependent in either a linear or
non-linear way although here we consider the linear schedule
$\beta(t)= \beta_0 (1 - t / T)$, as shown in
\autoref{fig:algorithmintuition}(b). Hence, the annealing term decreases to zero over time and the momentum
term vanishes for the equilibrium solution, which means that the \AIM\
algorithm finds a solution corresponding to the minimum of the
objective function $F(\bf{x})$ at time iteration $T$. Having such explicit stopping criteria is a lucid advantage of the
\AIM\ algorithm for an all-analog hardware implementation, as it
avoids the complexity of multiple intermediate readouts that
stochastic heuristic approaches suffer from \autocite{reifenstein2021coherent}.

We design a two-phase approach for the \AIM\ solver to operate similar to a black-box
solver that can quickly adjust the critical parameters within the given time limit,
as shown in \autoref{fig:algorithmintuition}(d). During the `exploration' phase, we evaluate the relative performance of \AIM\
algorithm across a vast range of parameters $(\alpha_0, \beta_0)$, where $\alpha_0$
scales the $\alpha$ parameter by the largest eigenvalue of the weight
matrix $Q$. A subset of `good' parameters is then passed for more extensive
investigation in the `deep search' phase (see Methods for details).

\section*{\AIM Hardware Design}
To surpass the speed of classical computers in solving optimization problems,
the \AIM\ algorithm is engineered to be amenable for analog hardware implementation. As a proof-of-concept, we design and build a fully analog \AIM\ architecture with discrete
opto-electronic off-the-shelf components widely available and common in telecommunication applications. We implement the most computationally expensive operation of the iterative update rule in
\autoref{eq:aim_iterative_update_rule}, namely the matrix-vector multiplication,
in optical domain and the annealing characterized by $\beta(t)$ in electrical domain.
The remaining momentum term could be implemented using capacitor-based time derivative circuits
in the electronic domain.

\autoref{fig:figure2}(a) describes the operational principles of the opto-electronic hardware.
In the optical domain, the values ${\bf x}_t$ are encoded in the intensity of light sources.
Each matrix element $Q_{ij}$, that represents the information about the
input problem, is encoded as the transmissivity of a single cell of an optical matrix of
modulators. In our specific implementation, we use multiple wavelength selective switches
to emulate this modulator matrix, with each one representing a row of the matrix.
As we show schematically in \autoref{fig:figure2}(a), the calculation of the
matrix-vector product in hardware is described by three steps. During the `fan-out' step, each
light source within ${\bf x}_t$ is split into $N$ optical replicas, where $N$ is the
size of the problem, resulting in $X_t$ matrix. In the `element-wise multiplication' step,
$Q_{ij} * (X_t)_{ij}$ is computed by shining each optical signal $(X_t)_{ij}$ onto an element of the array
of modulators with transmissivity $Q_{ij}$. During the last `fan-in' step, the
intensity-modulated optical signals are summed column-wise by collecting the light onto
$N$ photodetectors. As a result, the variables have moved onto the electrical domain in the
form of currents (voltages) which are proportional to the total optical power they have
received and, hence, to the $Q {\bf x}_t$ product. In such architecture, all the scalar
multiplications and additions involved in a single matrix-vector multiplication are computed
in parallel in a single pass of the optical sources through the discussed setup.
\begin{figure}
    \caption{\bodyfigurelabel{fig:figure2} \textbf{Hardware AIM setup and performance.} (a) The setup shows the experimental implementation
      of the \AIM\ algorithm based on opto-electronic components.
      The enhanced block represents the setup
      for variable $1$ with the two arms corresponding to the $\alpha$ and
      $\beta$ terms in \autoref{eq:aim_iterative_update_rule}. The
      all-to-all connectivity is realised in hardware through the
      wavelength selective switches, represented as `combiner' block
      in the scheme. (b) The signal intensities (green) and the corresponding objective
      optimality gap (blue) are shown as a function of time iterations
      for the $7$-bit \QUMO\ problem with $4$ binary and $3$ continuous
      variables. The filled light green and blue
      areas represent regions for steady state detection. The expected
      variable values and zero optimality gap are shown with grey dashed
      lines. (c) The success rate performance results are demonstrated for
      the hardware solver, its noisy simulated version, and
      noiseless ideal simulations of \AIM\ algorithm across $7$-bit
      \QUMO\ and \QUBO\ problems.
      (d) The transaction settlement problem instance involving three
      financial parties is mapped to the \QUMO\ abstraction and solved
      in hardware with a success rate of $100\%$. }
    \centering
\includegraphics[width=13.2cm]{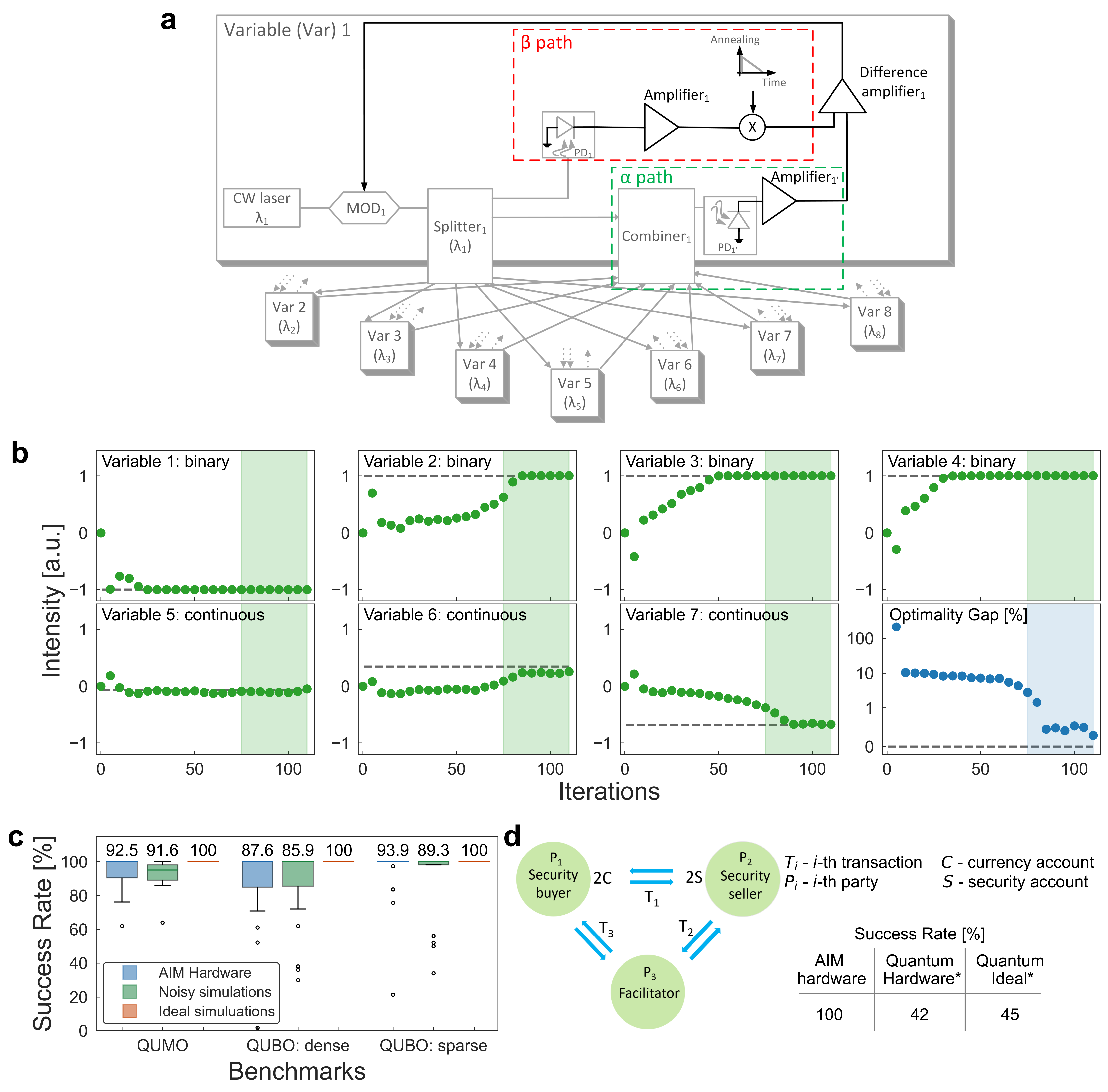}
\end{figure}

To implement the annealing, we use an additional replica of the signal ${\bf x}_t$ that is
optically path-matched with the objective gradient arm and captured with photodetectors. The annealing $\beta(t)$ path is realized with an analog mixer, in which the signal from the
path $\beta_0 {\bf x}_t$ is multiplied with a ramp signal that has negative slope. These electrical signals are further combined with the output signals from
the $\alpha$ path and supplied back to the corresponding modulated
sources, completing one iteration of the \AIM algorithm.

On each iteration, the nonlinearity is introduced into the system through the combination of
optical sources with the square-law detection scheme and the saturation of analog electronics.
By calibrating the gain of each amplifier prior to the modulator, we select a specific region
of the transfer function to realise either binary or continuous variables.
The closed loop system allows signals to remain in analog
domain for hundreds of iterations until the annealing term vanishes. At this point,
the signal amplitudes are read out digitally and the optimised objective value can be computed
for the original problem.

Opposed to the variable encoding in phase domain, encoding variables as light intensities can
bring a notable scaling advantage, as the precision requirements for calibrating variable paths
are proportional to the system ${\rm GHz}$ bandwidth and not the bandwidth of the optical sources,
which can reach $10^5 \ {\rm GHz}$. At the same time, the intensity encoding constrains one to operate with nonnegative matrix weights. For a general matrix-vector product, one can apply offsetting and scaling procedures to
represent operations with arbitrary matrix weights in analog
domain \autocite{goodman1978fully,wang2022optical}.

Such opto-electronic architecture offers a simple blueprint of analog hardware for
solving hard optimization problems with binary and continuous variables.

\section*{\AIM hardware performance and cross-validation}
We conduct a comprehensive evaluation of the opto-electronic \AIM\ solver on a diverse set of
\QUMO\ and \QUBO problems. The \AIM\ can solve optimization problems with binary and continuous
variables, with arbitrary connectivity, dense or sparse, and with $7$ bits accuracy for the
problem input weights. We vary the parameters $(\alpha_0, \beta_0)$ within their accessible range in hardware to
identify the optimal parameter combination for sampling. For a representative \QUMO instance, we present the time evolution of variables and optimality
gap in \autoref{fig:figure2}(b). The optimality gap is defined in a conventional way as the ratio of the difference between the
best objective and the found objective to the best objective. The steady states are detected within the shaded regions. For the measured binary and continuous variable values, we obtain a high-quality solution to the
\QUMO\ problem, with a final optimality gap of $0.29\%$, in an entirely analog manner without
any digital pre- and post-processing.

We cross-validate the simulated \AIM solver and its discrete opto-electronic implementation
on $50$ instances of the \QUMO\ and \QUBO\ models. Given the historical emphasis on the \QUBO\ model, we consider $40$ \QUBO instances with $8$
variables, representing problem classes that are known to be computationally challenging at scale. These instances are divided equally between two graph topologies, namely dense fully-connected
and sparse three-regular graphs.
In both cases, the matrix weight elements are drawn from the Gaussian
distribution and their bit precision is reduced to $7$ bits, resulting in instances belonging to
the Sherrington-Kirkpatrick \autocite{Kirkpatrick1975} and weighted maximum cut problems. Given the novel nature of the \QUMO\ abstraction, there is a lack of research on methods for
generating challenging small size instances with a mixture of binary and continuous variables.
We develop a technique for planting random continuous minimizer values in the global
solution and generate $10$ \QUMO instances with $7$ variables.
To make these instances more challenging to solve and obscure the planted
solutions, hundreds of random perturbations are applied to their matrix weights.

In \autoref{fig:figure2}(c), the success rates of the opto-electronic solver are presented for the
\QUBO\ and \QUMO\ instances. The success rate, a widely used metric for evaluating the performance of heuristic solvers, is
defined as the probability of finding the global objective value for a given set of parameters. The hardware solver targets exact objective values for the \QUBO\ instances while the relative
and absolute tolerances are set to $99.5\%$ for \QUMO\ instances to account for hardware
imperfections. For all benchmarks, the opto-electronic \AIM\ achieves high average success rates over $87.6\%$ with a
median success rate of $100\%$. This performance is in agreement with that of the simulated noisy \AIM\ algorithm, and is closely
aligned with the performance of the simulated noiseless \AIM\ algorithm.

The opto-electronic \AIM\ demonstrates exceptional performance on challenging synthetic \QUMO\
and \QUBO\ instances at small scale. While the hardware solver finds global objectives for all instances, there are several
outliers for which the success rates are lower, with values of about $62\%$, $1.7\%$, and $21.4\%$
across the \QUMO, \QUBO\ (dense), and \QUBO\ (sparse) benchmarks. Such outliers may be attributed to the inaccessible range of optimal parameter values or to
the particular nonlinear function implemented in hardware.

\section*{Solving Transaction Settlement Problem with \AIM hardware}

As a demonstration of the \QUMO\ abstraction importance, we further consider an industrially
important problem from financial domain, namely the transaction settlement problem, and
solve it with \AIM\ hardware.
This problem arises in the context of financial transactions when one needs to ensure
that the parties involved in the transaction are able to receive the funds or assets that
they are entitled to in a timely and efficient manner, as shown schematically in
\autoref{fig:figure2}(d). If one party fails to meet their obligations, the securities
settlement system may not be able to settle all transactions, which can cause a chain
reaction of many unsettled transactions. Consequently, transaction settlement is the
NP-hard optimization problem of finding the optimal set of transactions to settle
\autocite{gedin2020securities}.

Various approaches have been proposed for solving the transaction settlement problem, one
of which formulates it as a linear binary optimization problem with linear inequality
constraints \autocite{braine2021quantum}. In turn, the transaction settlement problem can be further mapped to the
\QUMO\ abstraction since the inequality constraints can be efficiently incorporated into
the objective function by introducing continuous slack variables.

Here we consider a problem instance that is derived from real settlement data
\autocite{braine2021quantum}.
In order to fit the current hardware requirements, we reduce the original small-scale
problem with $9$ variables to $6$ variables, i.e., $3$ binary and $3$ continuous. We
observe the high success rates of $99\%$ and $100\%$ for both, the hardware \AIM\ and
its simulated version in \autoref{fig:figure2}. For comparison, we highlight the
quantum hardware and simulated quantum hardware performance for the same problem
with a success rate of $45\%$ and $42\%$, respectively \autocite{braine2021quantum}.

\section*{Simulated \AIM Solver Performance at scale}

Here we validate the algorithmic performance of the \AIM\ solver that is implemented on the GPU. We consider a comprehensive set of benchmarks, one of which is the quadratic programming
library (QPLIB). The QPLIB benchmark is a collection of challenging synthetic and
real-world problems, gathered over a year-long open call from mathematical and numerical
analysis communities \autocite{QPLib:Paper,QPLib:Web}. For evaluating the \AIM\ solver on \QUMO\ problems, we focus on a subset of QPLIB benchmark,
namely the non-convex problems with linear inequality constraints. These problems have up
to several thousand variables, while the number of constraints reaches ten thousand. These
constraints lead up to ten thousand additional continuous variables for problems formulated
within the \QUMO\ abstraction. We consider unconstrained and equality-constrained binary
problems within QPLIB. In addition, we use the well-studied G-Set benchmark with synthetically generated problems
up to 20000 variables \autocite{Gset:Web}, Tile3D and Wishart instances from a recently
introduced CHOOK generator \autocite{perera2020chook}, as well as a set of manufacturing
problems, i.e., RCDP.

For \QUMO\ problems, we compare the \AIM\ approach against the commercial Gurobi solver,
that outperforms such optimization packages as Octeract, Baron, and Scip
\autocite{MittelmannBenchmarks:web}. For \QUBO\ benchmarks, in addition to Gurobi, we consider two
heuristic approaches, namely simulated annealing and parallel tempering, that are known
for a consistently better or similar performance over other physics-inspired methods
\autocite{aramon2019physics,mohseni2021nonequilibrium,tasseff2022emerging}. Both heuristic approaches benefit from highly-optimised implementations in Azure quantum
inspired optimization service and work as black-box solvers for a given time limit
\autocite{QIOservice:Web}.

We design the \AIM\ approach to resemble a black-box solver operation with the main sensitive
parameters $(\alpha_0, \beta_0)$ dynamically adjusted for a given time limit for each problem,
while small variations of the momentum parameter values are explicitly mentioned within and
between benchmarks (see Suppl. Mat.). For a representative performance comparison across such a disparate set of benchmarks,
we consider the quality of solution improvement metric:
\begin{align}
    \parbox{6em}{Objective\\ Improvement} =
\begin{cases}
    100 \% \cdot \frac{{\rm F_{AIM} \ - \ F_{Best \ rest}}}
                      {{\rm F_{Best \ known} \ - \ F_{Best \ rest}}}, & \parbox{13em}{if \AIM is
                        better than all\\ competing methods}\\
                      \\
                      0 \%,              & \parbox{13em}{if \AIM is equal to the best\\ competing method} \\
                      \\
    - 100 \% \cdot \frac{{\rm F_{Best \ rest} \ - \ F_{AIM}}}
                      {{\rm F_{Best \ known} \ - \ F_{AIM}}}, & \parbox{13em}{if \AIM is worse than
                      the best competing method}
\end{cases}
\end{align}
where ${\rm F_{Best \ known}}$ represents the best known minimum of the objective function
for the problem and the \AIM\ algorithm is compared against the best solution found by
competing solvers: Gurobi, parallel tempering, or simulated annealing.
Since all solvers are given the same computational resources equivalent to about $100$ seconds
of the \AIM solver, the solution improvement metric serves as a good indicator of their
relative performance in terms of finding better objective function values
(see Methods).

For the \QUMO\ benchmark, we report the \AIM\ speed-up against the Gurobi solver on the hardest
quadratic binary problems with linear inequality constraints within QPLIB (QPLIB:QBL). Since
Gurobi attempts not only to find the optimal solution, but also to prove the global optimality
of the solution, we consider the Gurobi time when it first finds the best objective value, which
can be compared with the time of the \AIM\ solver, that provides no global optimization
guarantees. In \autoref{fig:figure1}(a-b), we consider ten of the most difficult instances
requiring more than a minute of computational time for Gurobi to find the best known solution.
The one minute threshold is chosen as the problems that can be solved faster can be seen either
intrinsically simple or their structure could be substantially simplified by the
pre-processing techniques of Gurobi. The \AIM\ solver is up to three orders of magnitude faster in all \QUMO\ except the two instances,
one of which it is unable to solve. Moreover, the \AIM\ solver finds the new best solutions for two heavily
constrained instances in about $40$ seconds: the instances $3584$ and $3860$ have about $500$
binary and $10000$ continuous variables in \QUMO\ formulation. To evaluate the speed-up, we run
Gurobi for these two instances for five days. For the instance $3584$, Gurobi finds the same solution as the \AIM\ solver in about $54000$
seconds, while proving its global optimality takes four and half days. The Gurobi solver
optimises instance $3860$ to the same quality as the \AIM\ approach in about $13000$
seconds with an optimality gap of $8\%$ in five days.
\begin{figure}
    \caption{\bodyfigurelabel{fig:figure1} \textbf{Simulated \AIM\ solver performance.} (a) The relative speed-up performance of the \AIM\ solver compared
      to the Gurobi solver is shown on \QUMO\ instances formulated for
      a subset of QPLIB benchmark, namely instances with quadratic
      objective and linear inequality constraints for binary variables
      (QBL). The \AIM\ solver is up to three orders of magnitude faster on eight out of
      ten instances, and is unable to solve one instance. For
      instances $3860$ and $3584$, the \AIM\ solver finds the new state-of-the-art
      solutions. (b) The time performance is depicted for the Gurobi solver on
      the corresponding ten QBL instances in their original
      formulation.  These are the hardest instances within the QBL
      instances that take more than $60$ seconds for the Gurobi solver
      to find the best known solution. (c) The violin plots demonstrate distribution of the quality
      improvement performance for the \AIM\ solver compared to
      the best solution found by competing methods across six \QUBO\
      benchmarks. The competing methods include parallel tempering,
      simulated annealing, and Gurobi solvers.}
    \centering
    \includegraphics[width=8.6cm]{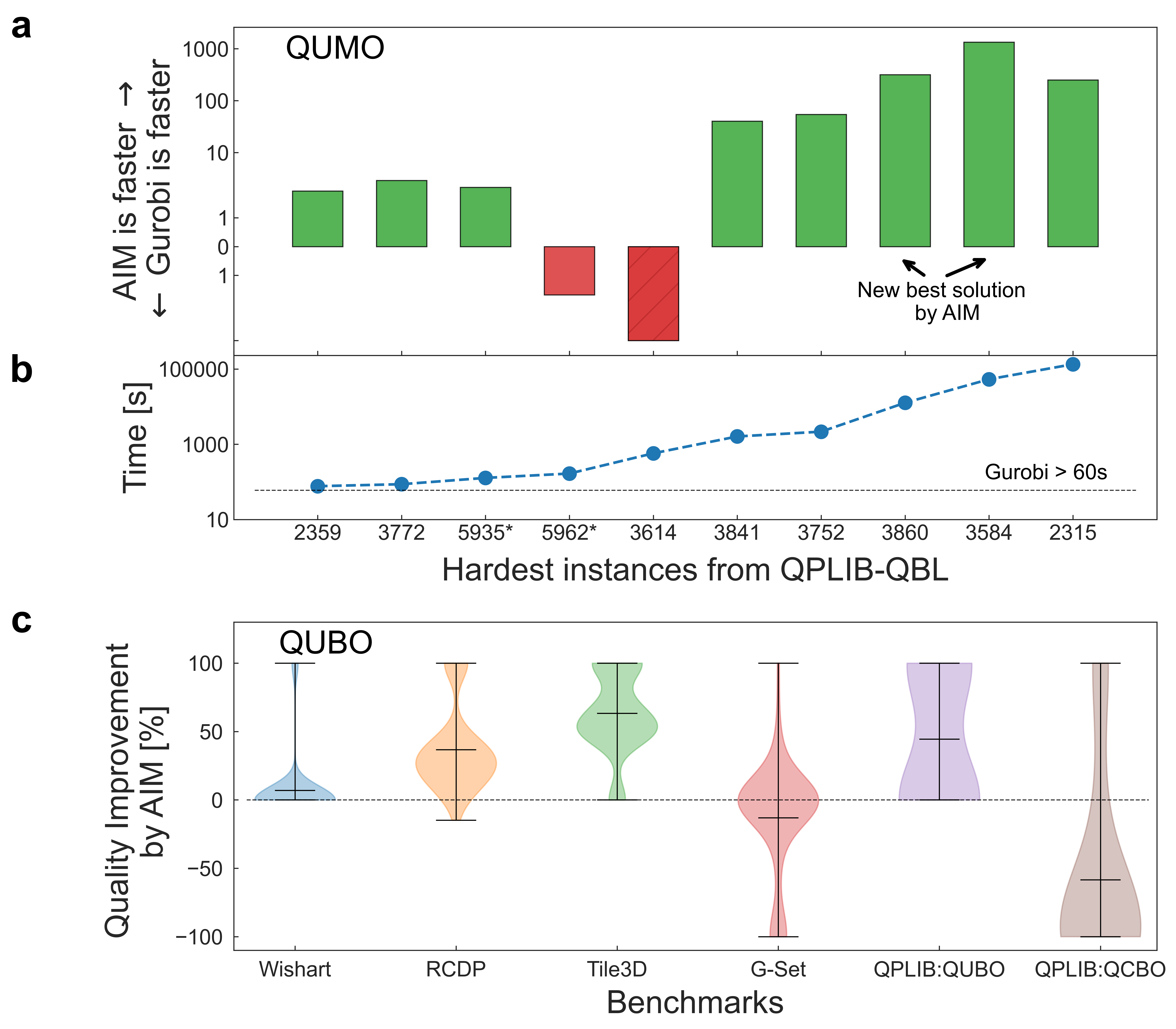}
\end{figure}

Although we do not know the origin of the hardest instances discussed above, we notice that
the ones solved by the \AIM approach are united by the same type of inequality
constraints, which can be seen as Horn clauses \autocite{hooker1999mixed}. The Horn clauses play fundamental role in
automated theorem proving and, intuitively, imply that if all variables except
one are true, then that last variable must also be true. Such insight into the logical
properties of this subset of the hardest QPLIB problems may imply that the \AIM solver can
efficiently propagate the logical clauses of Horn type that can bring advantageous performance
in solving logic programming problems.

In \autoref{fig:figure1}(c), the \AIM solver demonstrates consistent objective improvement
over the competing methods across several \QUBO benchmarks including
Wishart, RCDP, Tile3D, and QPLIB:QUBO, where QPLIB:QUBO includes a subset of graphs from QPLIB
that are natively represented as the \QUBO model. Remarkably, \AIM solver finds new best
solutions for the two largest problems with over thousand variables from QPLIB:QUBO, namely
$3693$ and $3850$ instances. For Gurobi, the optimization of instance $3850$ takes $110000$ seconds
to reach the same solution quality as \AIM solver in $40$ seconds, with the gap of $3\%$ in $5$
days. The instance $3693$ cannot be optimised to the similar \AIM quality of solution by Gurobi
in $5$ days. The \AIM solver further demonstrates competitive performance on the G-Set instances.
With many heuristic approaches applied to the G-Set benchmark over the past two decades, the
performance of the Gurobi solver has probably been overlooked due to the general perception that
it is inefficient at solving natively formulated \QUBO problems. In our analysis, we observe
that Gurobi solves particularly well instances with certain graph topologies, finding new best
global solutions for four graphs with torus geometry from G-Set benchmark, although it still
underperforms on other \QUBO\ instances.

Among all \QUBO\ benchmarks, \AIM\ solver is behind only on the quadratic constrained binary
optimization problems within QPLIB (QPLIB:QCBO), which includes instances with binary variables
and equality constraints. We note that Gurobi employs various pre-processing techniques that
can drastically reduce the number of variables and constraints \autocite{achterberg2020presolve},
thereby greatly simplifying problems before invoking the main optimization methods such as
simplex, interior point, and branch-and-bound.

With the development of similar pre-processing
techniques for the \AIM solver, the \QUBO\ and \QUMO\ formulations of constrained problems could be
simplified and one could expect an improved performance for the \AIM\ approach. For example, we
implement one of the pre-processing methods for solving two \QUMO\ instances within
QPLIB:QBL benchmark (see Methods).

\section*{Scalability Roadmap}

The developed opto-electronic \AIM\ architecture is the first tangible demonstration of
a fully analog, spatially-multiplexed hardware that is able to solve optimization problems
containing both binary and continuous variables with all-to-all connectivity. To address practical optimization problems at scale and guarantee the
speed-up improvement, we envision the same proposed \AIM architecture with $10000$ variables
using miniaturized and mass-produced opto-electronic components used, for example, in smartphone
devices. For instance, the transaction settlement problem can be split into multiple sub-problems
of that size based on transaction values and available liquidity. Another practical application at this scale could be the constrained portfolio optimization
problem. For this problem, the whole universe of stocks, e.g. $2800$ stocks within the New
York Stock Exchange, can be represented within the $10000$ variable hardware with
all-to-all connectivity and mixed variables.

The \AIM\ architecture with $10000$ variables will be achieved by
moving from discrete optical components to integrated opto-electronic
technologies from the consumer space that are low-cost, low-power and
scalable. Following the established operational principles of the
\AIM\ prototype, the most computationally expensive matrix-vector
product will be still implemented in optical domain, while the
annealing and momentum techniques will be realised in electrical
domain. For generating and detecting optical signals, i.e., problem
variables, one can use inexpensive and scalable micro-LEDs and CMOS
sensors.
To achieve sufficient system signal-to-noise ratios and dynamic ranges, each variable will
be represented by multiple dependent optical sources detected by multiple dependent photodetectors. Spatial light modulators (SLMs) could be a promising
technology for achieving high resolution modulator matrices,
with potentially each individual SLM pixel representing a different problem weight.
Estimating a source-to-SLM-pixel ratio of $1:100$, micro-LEDs
will shine onto rows of the modulators matrix to achieve the
all-to-all connectivity.
The modulated signals after the SLMs will be
collected column-wise onto CMOS sensors to perform massive dot-product
operations in parallel in a single pass of light through the system,
as in \autoref{fig:intro}. These illuminations, elongated in one dimension, can be implemented
with spherical-lens systems, which are well-corrected for errors in
large-field-of-view imaging \autocite{Nature/Comm/2022/Wang:OnePhoton}.

Assuming that each SLM pixel can provide the required element-wise
multiplication, an SLM resolution of $100$ million pixels is needed to
achieve a scale of tens of thousands variables.  Since a typical SLM
has several million pixels today, we envision the equivalent of a
two-dimensional array of SLMs with a resolution of about
$2000 \times 2000$ pixels each. We note that the current SLM refresh
rate of a few hundred Hz is sufficient as the \AIM\ solver needs to
compute millions of samples for each optimization problem. Such an
SLM, together with a $2000$ array of independent micro-LEDs and a $2000$ array of
independent CMOS sensors, would constitute the key building block, i.e., single `module' of
the \AIM\ solver. This module is then spatially multiplexed into an
array of modules whose input and output signals are properly split
and combined together guaranteeing a highly integrable platform for a
large matrix by vector multiplication.

Keeping into account the industrial technological advancements of micro-LEDs and
customized optical components, we calculate a total power consumption of the \AIM\
hardware of about $2 \ {\rm kW}$ at $10000$ variables, including the contributions of
transimpedance amplifiers and variable gain amplifiers. For the estimated
$9 \ {\rm cm}$ long overall path lengths of the whole hardware, we can
achieve a time per iteration of approximately $0.6 \ {\rm ns}$.

On the digital side, we run the GPU-based solver implementation on A100
over a large number of synthetic instances with $10000$ variables at FP16
precision. We measure the average time per iteration of $820 \ {\rm ns}$ at the
power consumption of $297 \ {\rm W}$. Assuming that the time per iteration
will decrease by a factor of two as we move to INT8 to fairly compare the simulated results
with the analog hardware, we calculate a speed up improvement of \AIM versus
state of the art digital hardware per unit power of about $100$ times at
$10000$ variables. With such improvement, the same optimization problem would be solved
in tens of seconds with the \AIM hardware, while an hour long computation would be required
for the digital solver.

 \section*{Discussion}

We face a pressing need for optimization hardware that can continue to scale in the edge
computing era. The computational complexity of optimization problems forces us to rely on
heuristics for finding approximate solutions. Heuristic algorithms running on a quantum computer
with a million physical qubits would be still four orders of magnitude slower than alternative
CPU-based algorithms \autocite{PRX/2021/Babbush:QuadraticSpeedups}. To address the shortcomings
of traditional classical and quantum
hardware, we harness the speed of light and aim to build an opto-electronic computer to solve hard
optimization problems at industrially-relevant scale more than two orders of magnitude
faster than state-of-the-art digital solutions.

Reflecting on the practical impact of existing unconventional approaches in optimization, we
focus on solving problems in the \QUMO abstraction. The \QUMO abstraction is superior to the
well-known \QUBO model, in which the objective variables are always binary, as it naturally
expresses a broader class of optimization problems. The advantage of the \QUMO abstraction is
caused not only by the cost of converting continuous variables to binary, but also by the
concomitant complexity increase for reformulated \QUBO problems of larger size.
On the contrary, \QUMO can represent the `sweet spot' between hardware amenability and higher
expressive power for many practical applications. We observe that the heavily-constrained
optimization problems, such as the transaction settlement or portfolio optimization problems,
have a rather straightforward and efficient transformation to the \QUMO abstraction.

To unleash the potential of analog computing, \AIM provides an original view at
unconventional optimization hardware. The \AIM algorithmic approach not only shows
highly-competitive performance across various benchmarks with four instances optimised to the
new best ever solutions, but also offers advanced gradient-based approaches that are easily
amenable for analog implementation. Following compute-in-memory principle, we utilize
commodity technologies to build the first \QUMO solver in analog hardware. The
small-scale opto-electronic \AIM based on discrete components solves the mixed binary and
continuous optimization problems with weight matrices accuracy up to $7$ bits in fully analog way. The
good quantitative agreement between the simulated solver and its optical counterpart paves a
promising avenue for realising the state-of-the-art optimization approaches in analog
hardware at scale.

We believe our blueprint for co-designing unconventional hardware and algorithms will ignite
the exploration of other optimization techniques and hardware platforms, as well as
enhance the development of automated problem mapping procedures to the \QUMO abstraction.
A wider variety of real world problems still need to be explored to understand which
applications would benefit the most from the \QUMO abstraction and for which the unconventional
hardware, such as \AIM, could bring a tangible advantage in terms of solution
quality and time-to-solution for reasonably large problems. At scale of tens of thousands
variables, \AIM could provide a route to more cost-effective, energy-efficient, fast and accurate
optimization architectures than conventional paradigms.

\begin{methods}

\section*{\AIM Parameters}

Typically, multiple hyperparameters need to be calibrated for heuristic methods
to achieve their best performance in solving optimization problems. To determine the optimal set of parameters, external hyperparameter optimization
packages or standard grid search techniques are widely used. While the choice of hyperparameters is critical, the time required to calibrate
them is often overlooked. However, the hyperparameter space grows exponentially with the number of parameters,
and a trade-off should be considered if the performance gain from fine-tuned additional
parameters outweighs the time spent for fine-tuning them. In principle, the three main parameters $\{\alpha, \beta_0, \gamma\}$ of the \AIM\ algorithm
need to be adjusted for each optimization problem. In our simulations, we notice that the algorithm is less sensitive to momentum
parameter value, while the $\alpha$ and $\beta_0$ values significantly affect the
solution quality. We further perform a linear stability analysis of the \AIM\ algorithm to evaluate
reasonable exploration regions for these two parameters and find that by scaling
the $\alpha$ parameter as $\alpha = \alpha_0 / \lambda$, where $\lambda$ is the
largest eigenvalue of the weight matrix $Q$, we get scaled parameters $\alpha_0$
and $\beta_0$ being in a similar optimal unit range across a wide range of problems.

We note that for two QPLIB:QUMO instances,
namely $5935$ and $5962$, we developed a pre-processing technique that greedily picks variables
with the highest impact on the objective functions and considers their possible values,
which is accounted in the reported time speed-up of \AIM.

The hardware performance is cross-validated with noisy simulations, in which the Gaussian
noise is injected into variables with standard deviations of $3\%$ for \QUMO problems
and $0.2\%$ for \QUBO problems to account for hardware imperfections.
In addition, the measured hardware nonlinearity function is used for realising binary variables
in the noisy simulations (further details are available in Suppl. materials). The smaller noise in case of \QUBO model could be attributed to the fact that the weights
of these instances are created to be sensitive to the noise in the first place.
The momentum term is zeroed for noisy simulations.
The relative and absolute tolerances are defined for some value $a$ and its reference value
$a_{ref}$ as $|a - a_{ref}| \leq ({\rm Tol}_{abs} + {\rm Tol}_{rel} |a_{ref}|)$.

\textit{Comment on objective improvement metric.}
The introduced objective improvement metric can be used evaluate the relative improvement
in objective value found by the solver of choice, i.e., the \AIM algorithm, compared to other
competing methods. In particular, the objective improvement of $100\%$ happens
when the \AIM\ approach finds the best known objective while the competing solvers cannot
achieve it, and in the reverse situation, the objective improvement is $-100\%$ when one of the
competing solvers finds the best known objective while the \AIM\ solver could not.

\section*{Competing solvers}
For a fair comparison, we ensure that all methods use similar computing resources.
Although the implementation of GPU or CPU based solvers can require highly varying
engineering efforts, we try to estimate the cost of running solvers on the hardware,
on which they are designed to run, and vary the time limit across solvers accordingly
to ensure similar cost per solver run.
In what follows, the \AIM solver runs on GV100 GPU for $5-300$ seconds per instance
across all benchmarks. For the highest time limit of \AIM\, we estimate time limits of
about $400$ seconds per instance for the simulated annealing and parallel tempering
methods run on multicore CPU machine. In case of Gurobi, our licence
allows one to use only up to $8$ cores, so it is given $1000$ seconds per instance.
To simplify the evaluations, the competing solvers are always given these maxed out
time limits even for the instances on which \AIM uses less than $300$ seconds.

\textbf{Simulated annealing.} As described in Azure quantum inspired optimization service
\autocite{QIOservice:Web}, simulated annealing is a Monte Carlo search type of method
that simulates a state of varying temperatures, where the temperature of a state
influences the decision-making probability.
For optimization problems,
the algorithm starts at an initial high-temperature state where ``bad" moves in the system
are accepted with a higher probability, and then slowly ``cools" on each sweep until the
state reaches the lowest specified temperature. At lower temperatures, moves that
don't improve the objective value are less likely to be accepted.
For \QUBO problems, each decision variable is ``flipped"
based on the objective value impact of that flip. Flips that improve the objective
value are accepted automatically. Flips that don't improve the objective value are
accepted on a probabilistic basis, calculated via the Metropolis Criterion.

\textbf{Parallel Tempering.} As described in Azure quantum inspired optimization service
\autocite{QIOservice:Web}, parallel tempering can be regarded as a variant of the
simulated annealing algorithm, or more generally Monte Carlo Markov Chain
methods \autocite{marinari1992simulated}.
As with simulated annealing, the cost function is explored through thermal jumps.
Unlike simulated annealing, a cooling temperature is not used. Instead of running a single
copy of the system, Parallel Tempering creates multiple copies of a system, called replicas,
that are randomly initialized and run at different temperatures. Then the same process is
followed as in simulated annealing, but based on a specific protocol two replicas can be
exchanged between different temperatures. This change can enable walkers that were
previously stuck in local optima to be bumped out of them, and thus encourages a
wider exploration of the problem space.

\textbf{Gurobi.} Gurobi is the commercial solver that is highly-optimized to work as
the black-box solver. Gurobi pre-solve techniques can drastically reduce the input problem size and the number
of constraints \autocite{achterberg2020presolve}.
We note that Gurobi finds for the first time the global minima
solutions to several largest G-Set instances including
G62, G72, G77, G81, which are all united by the same torus graph topology, and further proves
that the best known solutions are exact for other graphs with torus topology:
G11, G12, G13, G32, G33, G34, G48, G49, G50, G57, G65, G66, G67.

\section*{Benchmarks}
\label{sec:becnhmarks_discription}

\textbf{QPLIB benchmark.} The quadratic programming library (QPLIB) is a library
of quadratic programming instances \autocite{QPLib:Paper} collected over almost a year
long open call from various communities, with the selected instances being
challenging for state-of-the-art solvers. As described in the main part of the paper,
we consider only the hardest instances within the QPLIB:QBL class of problems, which contains
instances with quadratic objective and linear inequality constraints, the QPLIB:QCBO class
of problems which contains instances with quadratic objective and linear equality constraints,
and the QPLIB:QBN class of problems which contains \QUBO instances.

\textbf{Wishart benchmark.} Wishart planted ensemble (WPE) problems \autocite{wishart_paper}
are originally planted binary Integer Linear Programming (ILP) problems whose
coefficients are drawn from a correlated multivariate Gaussian distribution.
It has been shown that in the hard regime, the ground state is extremely difficult
to find using Monte-Carlo-based algorithms (such as parallel tempering) even for small
problem sizes of $32$ variables. The statistics has been collected across $100$ instances

\textbf{RCDP benchmark.} Rotationally constrained discrepancy problem (RCDP) arises in the
automotive manufacturing industry when one needs to arrange $n$ disks on a common axis
\autocite{RCDP_unpublished}. Due to imperfections, the disks have uneven surfaces and we wish
to decide the alignment of disks to
minimize total height. The disks have uneven
surfaces due to imperfect machining.
The goal is to rotate the disks to appropriate angles
with respect to a reference orientation such that when all put through the common axel,
the cumulative surface height in each sector is as close as possible to the ideal case
when all surfaces are perfectly flat. This problem can be formulated as either mixed integer
programming or QUBO. The RCDPs are tunable in hardness by increasing either $n$ and the number
of sectors $K$ or the correlation between the sectors. For our study,
the \QUBO problems were generated by external team. The
statistics has been collected across $100$ instances with $360$ variables.

\textbf{Tile3D benchmark.} The 3d tile planted problems \cite{tile3d_paper} are highly tunable
short-ranged Ising planted instances based on partitioning the problem graph into edge-disjoint
subgraphs. It has been shown that the tile-planted problems can be made orders of magnitude
(in terms of time-to-solution) harder than a typical 3D Gaussian spin-glass instance. The
statistics has been collected across $100$ instances with $512$ variables.

\textbf{G-Set benchmark.} The G-Set benchmark includes a collection of synthetically generated
instances from $800$ to $20000$ variables \autocite{ZIB/1997/37:Helmberg:SDP}.

\textbf{Hardware \QUBO instances.}
For the hardware experiments, we generate $7$-bit dense and sparse instances. The sparse
instances belong to the \QUBO\ model on three-regular graphs that is NP-hard
\autocite{garey1974some}, although NP-hardness
does not imply that every random instance is difficult to solve. We apply two additional procedures to make these instances more challenging to solve.
First, in order to ensure that the instances are not trivial to optimise, the global objective
minimizer is verified to be distinct from the projected eigenvector corresponding to the largest
eigenvalue of the weight matrix \autocite{kalinin2022computational}. Second, the instances are made highly sensitive to noise as we generate a million instances and
select ones that often lead to a different global objective solution in the presence of $1$-bit
level noise. For example, the generated instances with dense and sparse $7$-bit weight matrices are likely to
be affected by such small noise with a probability of $50-70\%$,
which may translate to success rates of up to $30-50\%$ for hardware that cannot guarantee
the target $7$-bit precision.

\textbf{Hardware \QUMO instances.}
Despite allowing several variables to have continuous values within the range of $[0, 1]$
in the \QUBO instances above, these variables tend to retain their binary values. To ensure that the given variables take indeed continuous values in the global
objective state, we plant random continuous minimizer values in the global
solution and generate $10$ \QUMO instances with $7$ variables. As the number of continuous variables increases for a given problem size, the problem instances
become relatively easier to solve. Consequently, we consider instances with one, two, and three
continuous variables.

\end{methods}

\noindent{\bfseries References}\setlength{\parskip}{12pt}\printbibliography[heading=none]

\begin{paddendum}
  \item[Author Contributions]
C.G., F.P. and H.B. conceived the project;
K.P.K. and C.G. developed the abstraction with the support of N.G.B. and H.B.;
K.P.K. and C.G. designed the algorithm with the support of N.G.B.;
K.P.K. and C.G. ran the simulations and analyzed the simulations data;
K.P.K. generated the QUMO and QUBO benchmarking used in the experiments;
F.P., G.M.-A., and I.H. designed the experiment; G.M.-A performed the experiments;
G.M.-A and K.P.K. analyzed the experimental data;
D.C. contributed technical advice and support on the electronic circuit design and implementation;
V.L. developed the automatized software for all the devices and equipments used in the experiments;
L.P. and J.C. contributed technical advice and ideas;
K.P.K., F.P., G.M.-A, C.G and H.B. wrote the manuscript;
A.R. supervised the project.
K.P.K. and G.M.-A. are the main contributors to this work and are listed in order of contributions.
The rest of the authors are listed alphabetically by last name.

  \item[Acknowledgments] The authors wish to thank Lee Braine, Barclays chief technology office,
  for useful discussions about optimization problems and \QUMO\ abstraction;
  Stephen Jordan, Brad Lackey, Amin Barzegar, Firas Hamze, Matthias Troyer and
  the MS Quantum team for useful discussions about optimization problems and
  for providing us with the \QUBO\ benchmarks (Wishart, Tile3D, RCDP).
  We acknowledge helpful discussions, encouragement, and support from our colleagues at
  Cloud Systems Futures at Microsoft Research, Cambridge, UK,
  and at M365 Research, Microsoft, Redmond, USA.
  N.G.B. thanks the Julian Schwinger Foundation grant JSF-19-02-0005 for the financial support.

  \item[Competing Interests] The authors of the paper have filed several patents relating to subject matter contained in this paper in the name of Microsoft Co.

  \item[Correspondence] Correspondence and requests for materials should be addressed
    to \emailstyle{project-aim-contact@microsoft.com}.

\end{paddendum}

\beginedfigures \newpage

\noindent{\bfseries \LARGE Supplementary Information}\setlength{\parskip}{12pt}

\section{Problem mapping advantage of \QUMO abstraction over \QUBO model}
\label{sec:qumo_mapping_advantage}

The realistic optimization problems are commonly constrained problems with mixed variable types.
The advantage of transforming an optimization problem to \QUMO abstraction rather than \QUBO
model is evident for a wide class of problems including such with both binary and continuous
variables, and problems with inequality constraints. To illustrate this mapping advantage, we
consider a toy quadratic optimization problem with a linear constraint formulated
in \autoref{fig:intro}:
\begin{eqnarray}
  {\rm Original \ problem: \quad} &\min& a x_1 x_2 + b x_3 \\
  &s.t.& 0 \leq c x_1 + d x_3 \leq 1,
\end{eqnarray}
in which all three variables can be assumed to be binary $x_i \in \{0, 1\}$. To get an
unconstrained optimization problem, the inequality constraint can be mapped to the objective
by using a penalty method:
\begin{equation}
  {\rm Unconstrained \ problem: \quad} \min a x_1 x_2 + b x_3 + P_0 (c x_1 + d x_3 + s - 1)^2,
\end{equation}
where $s \in [0, 1]$ is the continuous slack variable and $P_0$ is a large enough constant
that ensures that the constraint is satisfied. A \QUMO solver can be applied
directly to this unconstrained optimization problem. In contrast, if one has a \QUBO
solver, then the additional mapping step needs to take place. To map a continuous variable
to binary representation, one may consider either unary or binary encodings:
\begin{eqnarray}
  {\rm Unary \ encoding: \quad} s &=& \sum_{j = 1}^{2^{N_{bits}}} y_j \\
  {\rm Binary \ encoding: \quad} s &=& \sum_{k=0}^{N_{bits} - 1} 2^k y_k,
\end{eqnarray}
where $N_{bits}$ is the target bit precision for the continuous variable.
From this simple analysis, the problem mapping to the \QUMO abstraction is one-two orders
of magnitude more efficient than to \QUBO model in terms of the total number of variables.
For the hardware \QUBO solvers, the available bit
precision for the input problem weights needs to be taken into account, which may limit one
to use unary encoding.

\section{Comment on \QUMO abstraction}

The introduced \QUMO abstraction can be seen as a subclass of mixed integer nonlinear
programming (MINLP) class of problems. The MINLP problems appear in various fields
including  engineering design problems, particularly in chemical engineering where complex chemical
processes can be modelled using quadratic or other nonlinear functions, and integer
variables can represent discrete decisions. These optimization problems have a wide
range of applications in areas such as chemical process design and control,
network design, planning and scheduling, energy systems, and portfolio optimization.
To address these problems, a variety of tailored algorithms have been developed that
outperform general-purpose MINLP solvers. Given the increasing number of applications
for MINLP and the growing demand for powerful analytics and decision-making tools,
significant algorithmic developments are likely to emerge in this area over the next
decade, including advances in convexification, decomposition, and parallel implementations
to handle large-scale problems arising in machine learning.

Within MINLP, the mixed binary quadratic programming (MBQP)
class \autocite{brown2022copositive} would include constrained optimization problems
with binary and continuous variables, which is often also referred to as mixed
binary optimization (MBO) problems \autocite{QE/Lee/2021:MBOforFinance}.
Recent advancements in quantum and quantum-inspired technologies, as well as in optical
Ising solvers capable of approximately searching for the ground state of Ising spin
Hamiltonians, have increased interest in integrating Ising problems into the process of
solving difficult optimization problems. Existing approaches range from direct mapping
of MBQP to hybrid quantum-classical methods based on optimization algorithms
\autocite{brown2022copositive,QE/Lee/2021:MBOforFinance}.
The problems within MBQP class with linear constraints could
be efficiently translated to the \QUMO abstraction with an additional continuous
variable per each inequality constraint, making a wide range of applications more
accessible for solving directly on \AIM.

\section{Quantum hardware limitations for optimization problems}
\label{sec:quantum_hardware_limitations}

Quantum computing has emerged as a candidate hardware solver for hard optimization problems
\autocite{harrigan2021quantum}. Similar to the other physical machines, quantum computers also target the \QUBO\
abstraction. In addition to sharing the same abstraction limitations, additional quantum
hardware constraints further limit the potential of quantum computing as efficient solver for
optimization problems. Quantum approximate optimization algorithm (QAOA), that targets quantum
gate computers, performs similar to random guess on small-size \QUBO\ problems
\autocite{harrigan2021quantum} with theoretical estimations that even a million physical qubit
hardware will be still many orders of magnitude slower than existing classical heuristics
\autocite{sanders2020compilation}.
The quantum computers can offer up to quadratic speed-up over classical alternatives in
solving NP-hard optimisation problems, to which \QUBO and \QUMO belong. But, this quadratic
speed-up further suffers from the slow time operation of quantum gates.
Hence, unless new quantum optimization algorithms emerge and quantum computers scale
significantly, it is unlikely that quantum computing will allow us to tackle challenging
optimization problems at sizes of interest.

Quantum annealing platforms offer another approach for solving \QUBO problems.
The D-Wave pioneered \QUBO hardware solvers and managed to scale from tens of variables
to the current several thousands of variables over two decades \autocite{Dwave:Web}. In practice, one of the main
challenges for this hardware is the limited connectivity of only $15$ connections per each
variable, i.e. the Pegasus topology. This translates to additional mapping overhead of \QUBO
problem with an arbitrary topology to the D-Wave machine. In the worst case of the
fully-connected graph, the latest D-Wave Advantage hardware with $5000$ qubits can accommodate
only problem sizes up to $150$ variables. This one order of magnitude mapping overhead is
further amplified by one-two orders of magnitude mapping advantage of \QUMO over \QUBO abstraction.

\section{Examples of hardware time traces for \QUMO problems.}
\label{sec:hardware_variables_evolution}

The time evolution of variables and optimality gap are shown for \QUMO problems with one and two
continuous variables in \autoref{fig:qumo_hardware_time_trace_1conts} and
\autoref{fig:qumo_hardware_time_trace_2conts}, respectively. Initially, the feedback loop is open
in opto-electronic setup and all variables are set to zero. Once the feedback loop is closed,
the iterative update rule of the \AIM algorithm happens for about $100$ iterations and the
variables, i.e. signal intensities, evolve to their steady states. The final steady state has
near $0\%$ optimality gap and corresponds to the global minimum objective, as verified with the
Gurobi solver.

\begin{figure}[t]
  \includegraphics[width=17.2cm]{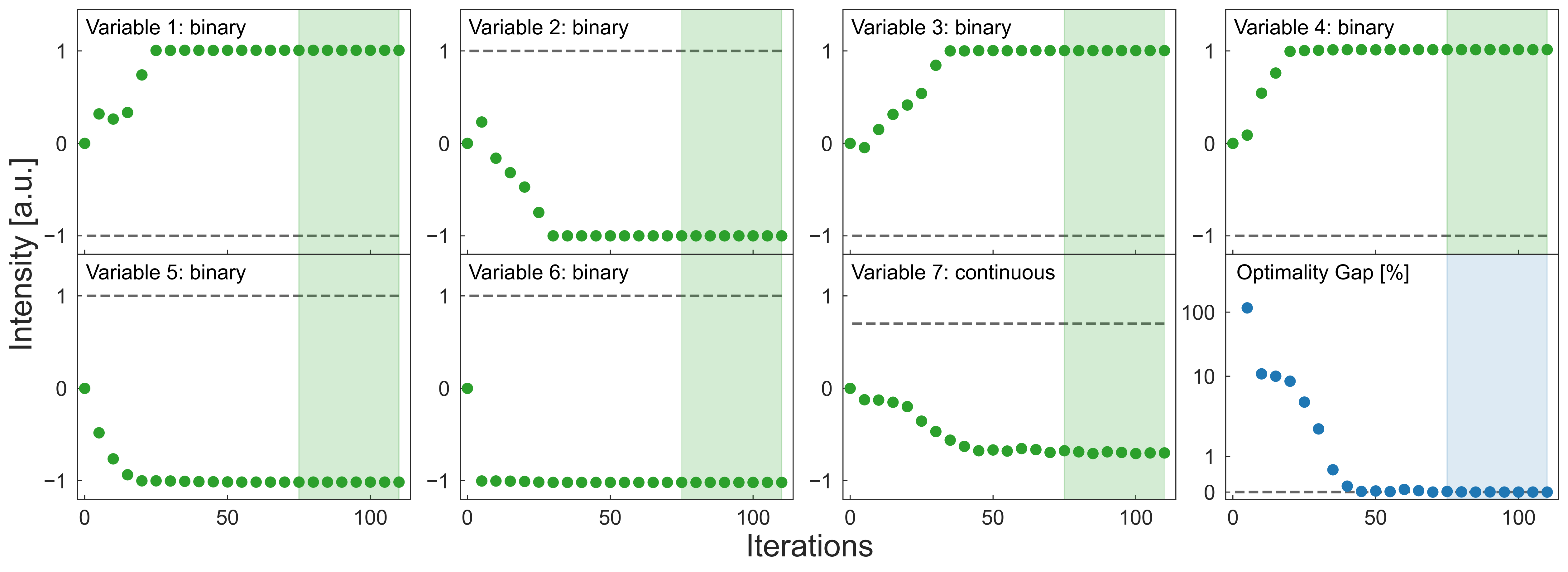}
  \caption{\edfigurelabel{fig:qumo_hardware_time_trace_1conts}
  \textbf{Hardware time traces for \QUMO problem with 1 continuous variable.}
  The signal intensities (green) and the corresponding objective optimality gap (blue) are shown
  as a function of time iterations for the $7$-bit \QUMO problem with $6$ binary and $1$
  continuous variables. The filled light green and blue areas represent regions for steady
  state detection. The expected variable values and zero optimality gap are shown with grey
  dashed lines. For the optimality gap graph, the symlog scale is used for y-axis for gap value
  over $1$ and linear scale for values below.}
  \centering
\end{figure}

\begin{figure}[t]
  \includegraphics[width=17.2cm]{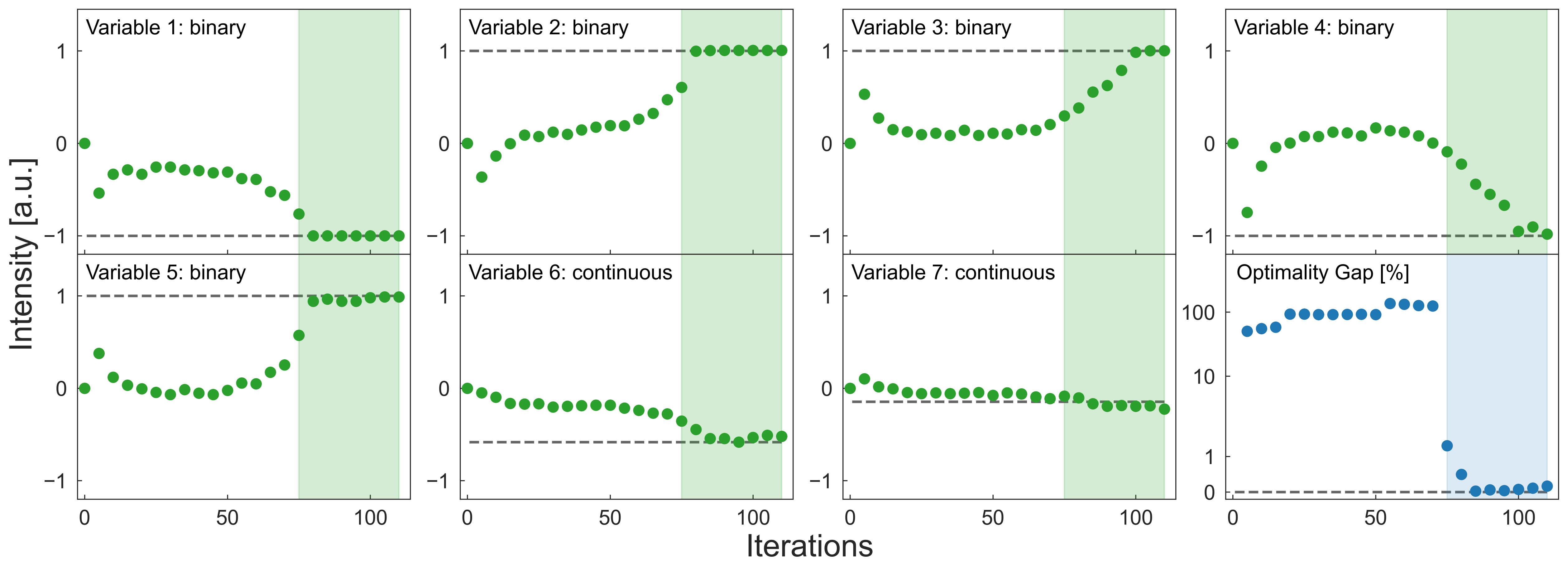}
  \caption{\edfigurelabel{fig:qumo_hardware_time_trace_2conts}
  \textbf{Hardware time traces for \QUMO problem with 2 continuous variable.}
  The signal intensities (green) and the corresponding objective optimality gap (blue) are shown
  as a function of time iterations for the $7$-bit \QUMO problem with $5$ binary and $2$
  continuous variables. The rest of description of \autoref{fig:qumo_hardware_time_trace_1conts}
  applies here as well.}
  \centering
\end{figure}

\section{Physical analogy of the \AIM algorithm}
\label{sec:physical_analogy_aim}
\AIM algorithm belongs to the family of gradient descent methods that use the concept of
momentum. The first momentum method has been introduced in 1964 by Boris Polyak
\autocite{Polyak/64:SpeedingUp} and is
known as heavy-ball method or simply momentum method. For optimising the nonlinear function
$F(x)$, it could be written as:
\begin{equation}
  {\bf x}_{t+1} = {{\bf x}_t} + \Delta t \left[ - \alpha \nabla F({\bf x}_t) +
  \gamma ({\bf x}_t - {\bf x}_{t - 1}) \right],
  \label{eq:heavy_ball_method}
\end{equation}
As discussed in the main text of the paper, the \AIM\ algorithm is
the advanced gradient descent method represented by the iterative update rule:
\begin{equation}
  {\bf x}_{t+1} = {{\bf x}_t} + \Delta t \left[ - \alpha \nabla F({\bf x}_t) - \beta(t) {\bf x}_t +
    \gamma ({\bf x}_t - {\bf x}_{t - 1}) \right].
    \label{eq:aim_iterative_update_rule_suppl}
\end{equation}
The momentum $\gamma$-term helps to overcome the issue of slow convergence due to fluctuations from
one iteration to the next, which can cause the state to `bounce' around an optimum instead of
continuously moving towards it. The momentum term reduces the `bouncing' by pushing consequent
updates in the direction of the most recent update \autocite{rumelhart1986learning}.
This effect provides a tendency for a system to continue along the path it is already taking,
therefore making it less susceptible to `bouncing' due to fluctuations in the gradient.
For example, if a variable equals to $1$ at iteration $i$ and is $-1$ on the iteration $i+1$,
then this variable will be pushed at the iteration $i+2$ by the momentum term for a value of
$2\gamma$, which provides a good local minima escape mechansim for the \AIM algorithm.
The parameter $\gamma$ is generally less than $1$ as the dependence of a given state on previous
states should weaken with the number of iterations between the current state and the earlier
state.

The physical interpretation for the momentum method \autoref{eq:heavy_ball_method} is well-known:
in terms of continuous dynamical system, the momentum parameter is equivalent to the point mass $m$
of Newtonian particles moving in viscous medium with friction coefficient $\mu$ under conservative
force field $f = \nabla E$ with potential energy $E(x)$ \autocite{qian1999momentum}:
\begin{equation}
  m \frac{d^2 {\bf x}}{dt^2} + \mu \frac{d {\bf x}}{dt} = - \nabla E( {\bf x}_t).
  \label{eq:newton_particle}
\end{equation}
Similarly, the iterative update rule in \autoref{eq:aim_iterative_update_rule_suppl} could be
interpreted as the the dynamical system of equations:
\begin{equation}
  m \frac{d^2 {\bf x}}{dt^2} + \mu \frac{d {\bf x}}{dt} = - \nabla F({\bf x}) - \phi(t) {\bf x},
  \label{eq:newton_particle_aim}
\end{equation}
where $\phi(t)$ is the nonconservative force. To find the relation between physical quantities and
parameters $\alpha$, $\beta$, and $\gamma$, one can discretise the
\autoref{eq:newton_particle_aim}:
\begin{equation}
  m \frac{{\bf x}_{t + \Delta t} - 2 {\bf x}_t + {\bf x}_{t - \Delta t}}{\Delta t^2} +
  \mu \frac{{\bf x}_{t + \Delta t} - {\bf x}_t}{\Delta t} = - \nabla F({\bf x}_t) - \phi(t) {\bf x}_t,
\end{equation}
which can be further rewritten as
\begin{equation}
  {\bf x}_{t + \Delta t} = {\bf x}_t - \frac{\Delta t^2}{m + \mu \Delta t} \nabla F({\bf x}_t)
  - \frac{\Delta t^2}{m + \mu \Delta t} \phi(t) {\bf x}_t
  + \frac{m}{m + \mu \Delta t} ({\bf x}_t - {\bf x}_{t - \Delta t}).
  \label{eq:newton_particle_aim_discretised}
\end{equation}
By comparing terms in \autoref{eq:aim_iterative_update_rule_suppl} and
\autoref{eq:newton_particle_aim_discretised}, we get:
\begin{eqnarray}
  \alpha \Delta t &=& \frac{\Delta t^2}{m + \mu \Delta t} \\
  \beta(t) \Delta t &=& \frac{\Delta t^2}{m + \mu \Delta t} \phi(t) \\
  \gamma \Delta t &=& \frac{m}{m + \mu \Delta t}.
\end{eqnarray}
Thus, the Newtonian system of equations, described by \autoref{eq:newton_particle_aim}, is
equivalent to the \AIM iterative update rule, described by
\autoref{eq:aim_iterative_update_rule_suppl}, with the parameters $\alpha$ representing the scaling
factor of the conservative force, $\beta(t)$ standing for the nonconservative force, and $\gamma$
corresponding to the particle mass.

The \AIM algorithm converges to steady states for any positive values $\alpha$, $\beta(t)$,
and $\gamma$. These steady states are the minima of the Lyapunov function, representing the
energy of the system.
Denoting $y = f_{\rm nonlinear}(x)$ to represent either binary or continuous variables, achieved
by applying the elementwise nonlinear function for binary variables and linear function for
continuous variables, the Lyapunov function can be written for the \QUMO objective $F(y)$ for
\autoref{eq:newton_particle_aim} as:
\begin{equation}
  E = \frac{m}{2} \frac{d f_{\rm nonlinear}^{-1}({\bf y}^T)}{dt} \frac{df_{\rm nonlinear}^{-1}({\bf y})}{dt}
  - \frac{1}{2} {\bf y}^T Q {\bf y} - b^T {\bf y} + \phi(t)
  \sum_{i = 1}^N \int_0^{y_i} f_{\rm nonlinear}^{-1}(x) dx.
  \label{eq:lyapunov_function_aim}
\end{equation}
Due to presence of nonconservative force, the overall improving \QUMO\ objective value during
the time evolution can be worsening for several intermediate consecutive iterations. Towards
the end of the time evolution, the annealing term $\phi(t)$ decreases to zero and the kinetic
term vanishes, so the minima of the Lyapunov function correspond to the equilibrium states of
the \autoref{eq:newton_particle_aim} and are the minima of the \QUMO objective function. From
the physics perspective, the convergence to the overall better optimal state in the end happens
as the energy decreases due to the presence of friction.

We note that the promising results of analog computations have been demonstrated for solving
differential equations and performing complex mathematical operations. For example, the
Fredholm integral equations of the second kind can be solved using free-space
visible radiation \autocite{cordaro2023solving},
the Hilbert transformation can be performed with a second-order
optical integrator \autocite{ashrafi2015time},
the basic trigonometric operations are enabled with
metasurface-based platform \autocite{zhao2022deep},
and mathematical operations can be performed using high-index
acoustic metamaterials \autocite{zangeneh2018performing}.

\section{Comparison of the \AIM architecture to other iterative approaches}
In case of the \QUMO objective function, the simulated iterative update rule of the \AIM
algorithm reads as:
\begin{equation}
  {\bf x}_{t+1} = {\bf x}_t + \Delta t \left[ \alpha (Q f_{\rm nonlinear}({\bf x}_t) + b^T)  -
  \beta(t) {\bf x}_t +
  \gamma ({\bf x}_t - {\bf x}_{t - 1}) \right],
\end{equation}
\[
  f_{\rm nonlinear}([{\bf x}_t]_i) =
\begin{cases}
    sign([{\bf x}_t]_i),& \text{if } [{\bf x}_t]_i - \text{binary}\\
    [{\bf x}_t]_i,      & \text{if } [{\bf x}_t]_i - \text{continuous},
\end{cases}
\]
where each variable $[{\bf x}_t]_i$ is also clipped to the range of $[-1, 1]$. For the well-studied
\QUBO model, the \AIM algorithm can be further rewritten in a simpler form as:
\begin{equation}
  {\bf x}_{t+1} = {\bf x}_t + \Delta t \left[ \alpha (Q \cdot sign({\bf x}_t) + b^T)  -
  \beta(t) {\bf x}_t +
  \gamma ({\bf x}_t - {\bf x}_{t - 1}) \right].
\end{equation}
where all variables are now assumed to be binary. We choose sign nonlinearity here, but other
nonlinearities could be considered including $\{ \cos, \tanh, clamp \}$. As we discussed above in
\autoref{sec:physical_analogy_aim}, the \AIM iterative update rule corresponds to the second
order differential equation. Consequently, the \AIM algorithm is distinct from all the
first order methods, the well-known example of which is the Hopfield networks
\autocite{PNAS/1982/Hopfield:Network,hopfield1985neural}. The Euler
update rule and ordinary differential equation can be written for Hopfield networks as:
\begin{eqnarray}
  {\bf x}_{t+1} &=& {\bf x}_t + \Delta t \left[ \alpha (Q \cdot tanh({\bf x}_t) + b^T)  -
  \beta {\bf x}_t \right], \\
  \frac{d{\bf x}}{dt} &=& \alpha (Q \cdot \tanh({\bf x}) + b^T)  - \beta {\bf x}.
\end{eqnarray}
In these equations, the losses term is represented by a constant parameter $\beta$.
By considering time-varying losses $\beta(t)$ in Hopfield networks, one arrives to the equations,
that are fundamentally similar to our current opto-electronic \AIM implementation.

Another recently introduced heuristic method is called the simulated bifurcation algorithm
\autocite{goto19optimization,goto2021high} which
is described by equations:
\begin{eqnarray}
  \frac{d {\bf x}}{dt} &=& a_0 {\bf y}, \label{eqs:toshiba_system1} \\
  \frac{d {\bf y}}{dt} &=& c_0 Q \cdot sign({\bf x}) - (a_0 - a(t)) {\bf x},
  \label{eqs:toshiba_system2}
\end{eqnarray}
where $a_0$, $c_0$, and $a(t)$ are the hyperparameters. This system of equations can be further
reduced to the single second order differential equation:
\begin{equation}
  \frac{d^2x}{dt^2} = c_0 Q \cdot sign(x) - (a_0 - a(t))x.
\end{equation}
Compared to the \AIM algorithm equations, the simulated bifurcation method doesn't have the first
order derivative and its iterative update rule can be written as:
\begin{equation}
  {\bf x}_{t+1} = {\bf x}_t + ({\bf x}_t - {\bf x}_{t-1}) + \Delta t^2 \left[ -
  (a_0 - a(t)){\bf x}_t + c_0 Q \cdot sign({\bf x}_t) \right].
\end{equation}
This update rule is similar to the \AIM algorithm with $\gamma = 1$. In general, the momentum
methods with $\gamma \geq 1$ are unstable \autocite{torii2002stability}. Such instability is probably mitigated
in the simulated bifurcation algorithm by an extra condition that manually forces the
difference $({\bf x}_t - {\bf x}_{t-1})$ to be zero if $|{\bf x}_t| = 1$. We also note that the original iterative
update rule of the simulated bifurcation algorithm uses symplectic Euler method for discretising
the system of equations above and can be written as:
\begin{equation}
  {\bf x}_{t+1} = {\bf x}_t + a_0 \Delta t \left[ ({\bf x}_t - {\bf x}_{t-1}) - \Delta t (a_0 - a(t)){\bf x}_t
            + \Delta t c_0 Q \cdot sign({\bf x}_t) \right].
\end{equation}
which would suffer from the same instability issue due to the momentum term without applying the
extra constraint above.

\section{Extensions of AIM algorithm}

\textbf{Extenstion 1: other momentum-based approaches for the \AIM algorithm.}
Momentum methods are commonly used in machine learning to accelerate the
training of neural networks. Besides the heavy-ball method, one may modify the \AIM algorithm
to be based on, for example, the Nesterov momentum method, in which case the update rule would
be given by the following equation:
\begin{equation}
  {\bf x}_{t+1} = {\bf x}_t + \Delta t \left[ - \alpha \nabla F({\bf x}_t + \gamma ({\bf x}_t  - {\bf x}_{t-1})) - \beta(t) {\bf x}_t
          + \gamma ({\bf x}_t - {\bf x}_{t-1}) \right].
\end{equation}
Unlike heavy-ball method, the Nesterov momentum update evaluates the gradient
at a point $({\bf x}_t + \gamma ({\bf x}_t  - {\bf x}_{t-1}))$, to which the momentum term has been applied,
instead of evaluating the gradient at the most recent state ${\bf x}_t$.

The performance comparison between \AIM algorithms based on different momentum-based methods is
a promising avenue for future studies.

\textbf{Extension 2: Exploration phase.}
In order to limit the range to be searched for each parameter, the initially defined bounds may be
further dynamically updated based on a computed performance of solutions within the bound in
comparison with solutions for parameters outside of the current bound. In other words, given a
current bounds for $\alpha \in [\alpha_{min}, \alpha_{max}]$, by increasing the bound to, for
example, $10*\alpha_{max}$ and evaluating solutions for parameters within the increased bound,
it may be determined that solutions outside the current bound are better than solutions inside
the current bound and in this case the bound is increased. This process may be repeated over
many iterations until a suitable region of parameter space is found.

\section{Equivalence of \QUBO, maximum cut, and Ising models}\label{sec:formulations}

The \QUBO, weighted maximum cut, and the problem of minimization of the classical Ising Hamiltonian
are mathematically equivalent. The \QUBO\ problem
seeks to find assignments to a set of binary variables $x \in \{0, 1\}$ with the goal of
minimizing the objective: \begin{equation}
  \min_x F_{\rm QUBO} = \min_{\bf x} - 0.5 {\bf x}^T A {\bf x} - {\bf a}^T {\bf x} - a_0
  \label{eq:qubo}
\end{equation}
where the matrix $A$, the vector $a$, and the constant $a_0$ are problem specific.

In the Ising model, one typically referes to variables as spins $y \in \{-1, 1\}$, to
quadratic weights as interactions $B$, and to linear terms as magnetic fields ${\bf b}$. The
objective is then to minimise the energy function:
\begin{equation}
  \min_y F_{\rm Ising} = - 0.5 {\bf y}^T B {\bf y} + {\bf b}^T {\bf y} + b_0.
  \label{eq:ising}
\end{equation}
Substituting ${\bf y} = 2 {\bf x} - 1$ into \autoref{eq:ising}, one gets the mapping between
the \QUBO and Ising models:
\begin{eqnarray}
  A &=& 4 B \\
  {\bf a} &=& 2 {\bf b} - 2 B {\bf e} \\
  a_0 &=& b_0 - {\bf b}^T {\bf e} + 0.5 {\bf e}^T B {\bf e}
\end{eqnarray}
where ${\bf e}$ is the unity vector.

In the weighted maximum cut model, one looks for the cut of the given graph $C$ into two parts with
maximised number of their connecting weighted edges:
\begin{equation}
  \max F_{\rm maxcut} = \max_{E_1, E_2} \sum_{i \in E_1 \\ j \in E_2} C_{ij}
  \label{eq:maxcut}
\end{equation}
where $E_1$ and $E_2$ represent vertices of two subgraphs. One can see the maximum cut model
is equivalent to the problem of minimization of the Ising Hamiltonian by noticing that:
\begin{equation}
  \max F_{\rm maxcut} = \max_y \sum_{i,j} C_{ij} \frac{1 - y_i y_j}{2} = \max_{\bf y} - 0.5 {\bf y}^T C {\bf y}
  + \sum_{i, j} C_{ij}.
\end{equation}
We further note that the linear terms could be incorporated within quadratic terms by
introducing an extra variable.

\end{document}